\documentclass[conference]{IEEEtran}

\usepackage{xcolor}

\usepackage{url}

\usepackage{booktabs} 
\usepackage{amsmath}
\usepackage{fixltx2e}
\usepackage{multirow}
\usepackage{xcolor}
\usepackage{gensymb}
\usepackage{listings}
\usepackage{nicefrac}
\usepackage{threeparttable}
\usepackage{array}
\usepackage[]{cite}
\usepackage{graphicx}
\usepackage{caption}
\usepackage{subcaption}
\usepackage{siunitx}
\usepackage{placeins}
\usepackage{romannum}

\usepackage{amssymb}
\usepackage{pifont}

\usepackage[mathlines,switch,modulo]{lineno}
\usepackage{color}

\usepackage{algorithm,algpseudocode}
\usepackage[font=small,labelfont=bf,
  justification=justified,
  format=plain]{caption}
\usepackage{dblfloatfix}

\usepackage[bottom,hang,flushmargin,para]{footmisc}

\newcommand{\ignore}[1]{}

\begin{document}

\bstctlcite{IEEEexample:BSTcontrol}

\title{Performance Implications of NoCs on 3D-Stacked Memories: Insights from the Hybrid Memory Cube}


\author{\IEEEauthorblockN{Ramyad Hadidi, Bahar Asgari, Jeffrey Young, Burhan Ahmad Mudassar, Kartikay Garg,\\Tushar Krishna, and Hyesoon Kim}
\IEEEauthorblockA{
Georgia Institute of Technology\\
Email: \{rhadidi,bahar.asgari,jyoung9,burhan.mudassar,kgarg40,\}@gatech.edu,\\ tushar@ece.gatech.edu,hyesoon@cc.gatech.edu}
\vspace{-0.0in}}

\maketitle
\pagenumbering{arabic}
\thispagestyle{plain}
\pagestyle{plain}

\begin{abstract}
Three-dimensional (3D)-stacked memories, such as Hybrid Memory Cube (HMC), provide a promising solution for overcoming the bandwidth wall between processors and memory by integrating memory and logic dies in a single stack. Such memories also utilize a network-on-chip (NoC) to connect their internal structural elements and to enable scalability. This novel usage of NoCs enables numerous benefits such as high bandwidth and memory-level parallelism and creates future possibilities for efficient processing-in-memory techniques. However, the implications of such NoC integration on the performance characteristics of 3D-stacked memories in terms of memory access latency and bandwidth have not been fully explored.
This paper addresses this knowledge gap (\romannum{1}) by characterizing an HMC prototype using Micron's AC-510 accelerator board and by revealing its access latency and bandwidth behaviors; and (\romannum{2}) by investigating the implications of such behaviors on system and software designs. Compared to traditional DDR-based memories, our examinations reveal the performance impacts of NoCs for current and future 3D-stacked memories and demonstrate how the packet-based protocol, internal queuing characteristics, traffic conditions, and other unique features of the HMC affects performance of applications.
\end{abstract}
\vspace{0pt}

\IEEEpeerreviewmaketitle

\section{Introduction}
In the past decade, the demand of data-intensive applications for high-performance memories has pushed academia and industry to develop novel memories with larger capacity, higher access bandwidth, and lower latency. To this end, JEDEC-based memories (i.e., DDRx) have evolved to include three-dimensional (3D)-stacked DRAMs, such as High Bandwidth Memory (HBM)~\cite{lee:kim14}. While such memories are compatible with traditional architectures and JEDEC standards, they are limited in terms of scalability and bandwidth which is due to their wide buses and the use of the standard DDRx protocol. Therefore, a generation of 3D-stacked memories with packet-based communication has been introduced and is currently implemented in the Hybrid Memory Cube, or HMC~\cite{jed:kee12}. Thanks in part to an internal packet-switched network and high-speed serial links between the processor and memory stack, this type of novel 3D-stacked memory exploits both internal and external networks to extend its capacity and scalability~\cite{paw11, kim:kim13}. The HMC consists of vertical memory partitions called vaults and a logic layer which consists of memory controllers (i.e., vault controllers) connected via an internal network-on-chip (NoC)~\cite{hybrid2013hybrid1}. As our analysis shows, the characteristics and contention of this internal NoC play an integral role in the overall performance of the HMC.

\begin{figure}[b]
\centering
\vspace{-0.25in}
\includegraphics[width=1\linewidth]{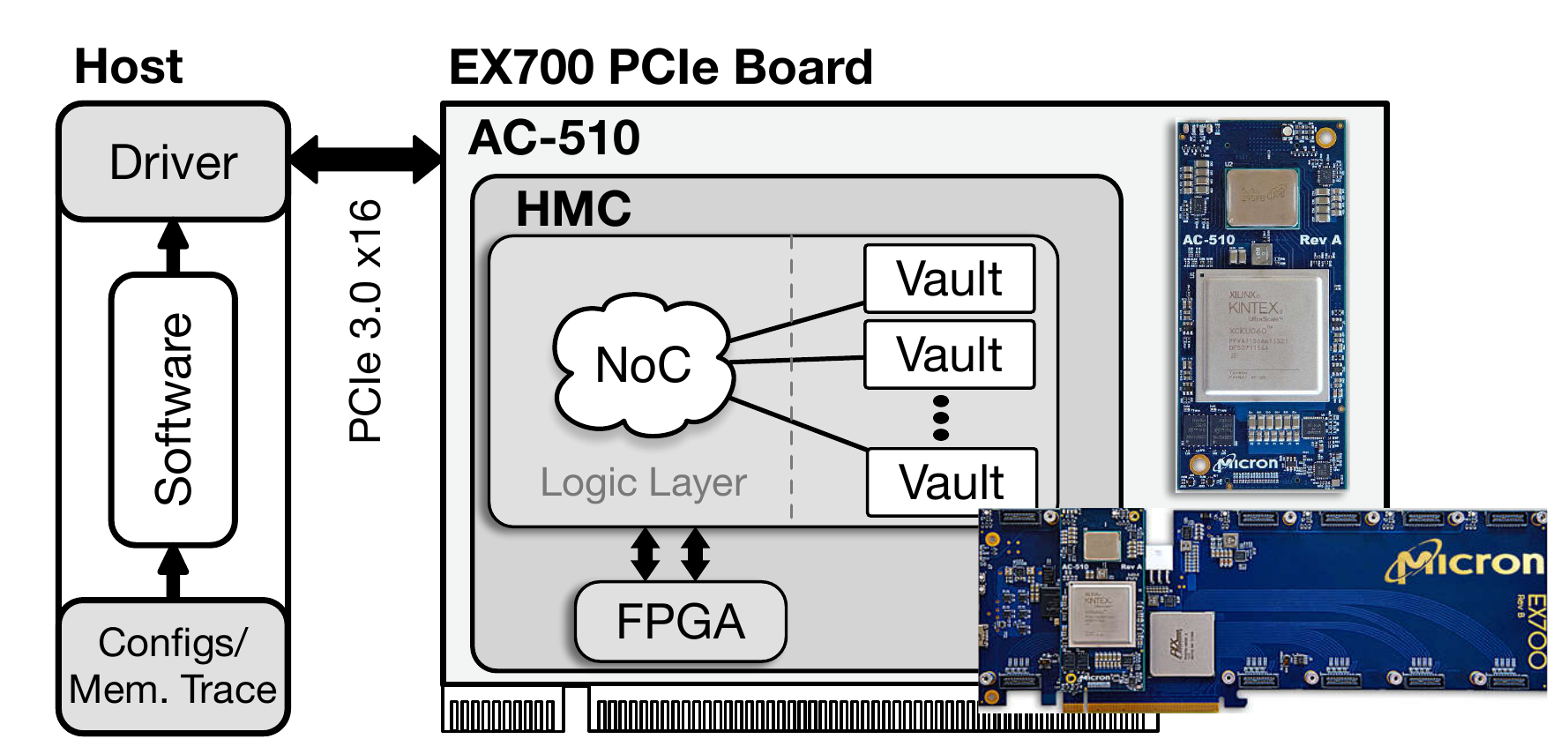}
\vspace{-0.0in}
\captionsetup{singlelinecheck=on,aboveskip=-8pt,belowskip=-3pt}
\caption{An overview of our system, and NoC of the HMC.}
\label{fig:intro}
\vspace{-0.05in}
\end{figure}

Logic and memory integration within 3D stacks has motivated researchers to explore novel processing-in-memory (PIM) concepts within the architecture of 3D-stacked memories using simulation~\cite{bla:ann06, kim:kim13, zhao:sun13, pug:jes14, zha:jay14, jun:sun:sun15, hsi:ebr16, nai:kim:2015, nai:had17, had:nai17}. However, few researchers have studied actual prototypes of memories similar to the HMC~\cite{gok:llo15, sch:fro2016, ibr:fat2016, had:asg17}. In particular, to the best of our knowledge, no experimental work has sought to characterize the bandwidth and latency\footnotemark\,impacts of the internal NoC on the performance of the HMC. In addition to understanding the performance impacts of this NoC on applications, such characterizations are also important for the design of PIM units built around or inside the HMC. In order to gain further insights into the impacts of the internal NoC on 3D-stacked memories, we evaluate the performance characteristics of an HMC\,1.1~\cite{hybrid2013hybrid1} prototype. We utilize a Xilinx FPGA and an HMC\,1.1 on the Micron's AC-510~\cite{ac510} accelerator board, which is mounted on an EX-700~\cite{ex700} PCIe backplane. Figure~\ref{fig:intro} presents the full-stack overview of our FPGA-based evaluation system, which includes user configurations, memory trace files, software, driver, an FPGA, and an HMC.

\footnotetext{Latency and round-trip time are interchangeably used in this paper.}

Our analyses characterize access properties for both low- and high-contention traffic conditions, for which we use two combinations of software and digital designs (i.e., Verilog implementations on the FPGA). Our results reveal (\romannum{1}) latency and bandwidth relationships across various access patterns, targeted to structural organizations of the HMC (i.e., vaults and banks), (\romannum{2}) latency distributions across the vaults of HMC , (\romannum{3}) quality of service (QoS) within a particular access pattern, and (\romannum{4}) bottlenecks to occur within the HMC, associated infrastructure, or within each access pattern. The contribution of this paper are the following:

\begin{itemize}
  \item This is the first study, to the best of our knowledge, that explores the bandwidth and latency impacts of the internal NoC of the HMC, a prototype of packet-switched 3D-stacked memories.
  \item It examines how the internal NoC behaves under low- and high-contention traffic conditions, presents the concept of QoS for 3D-stacked memories, and describes how future system and application designs should incorporate the HMC to achieve desirable performance.
  \item It presents a detailed analysis of the latency distribution that is caused by the internal NoC of the HMC for a specific access pattern and related consequences and opportunities.
  \item It studies request and response bandwidth relationships for various access patterns, determines the source of bottlenecks, and presents solutions for avoiding them. 
\end{itemize}

In the rest of this paper, we first review the HMC\,1.1 specification in Section~\ref{hmc} and then introduce our infrastructure and methodology in Section~\ref{sec:meth}. After that, Section~\ref{sec:res} presents and analyzes the details of latency and bandwidth of the HMC with various traffic conditions and the contribution of the NoC in each scenario. Subsequently, Sections~\ref{related} and \ref{sec:con} review related work and present conclusions based on our analyses.

\section{Background}
\label{hmc}

In this paper, we focus on the HMC\,1.1 specification (\emph{Gen2})~\cite{hybrid2013hybrid1}, currently available for purchase.  This section presents background on the HMC structure and relevant information on packet-based memories for our analyses.

\subsection{HMC Structure}
\label{sec:hmc-struc}
\noindent

\begin{figure}[b]
\centering
\vspace{-0.1in}
\includegraphics[width=0.9\linewidth]{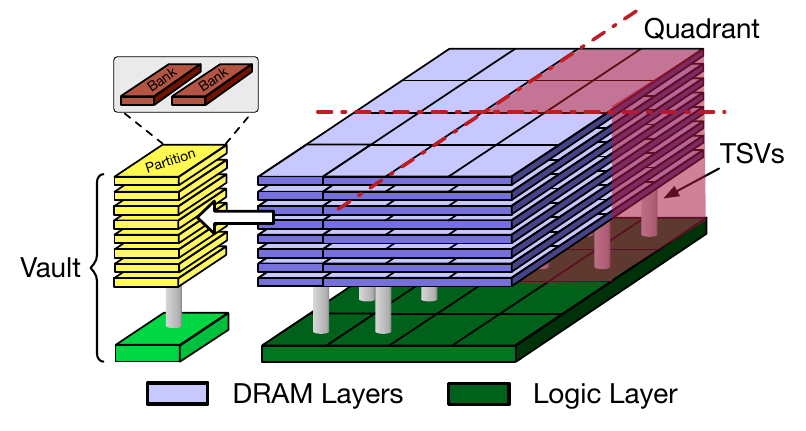}
\vspace{-0.0in}
\captionsetup{singlelinecheck=on,aboveskip=4pt,belowskip=-5pt}
\caption{4\,GB HMC\,1.1 internal structure.}
\label{fig:hmc-struc}
\vspace{-0.0in}
\end{figure}

The HMC\,1.1 consists of eight DRAM dies stacked on top of a logic die, vertically connected by 512 \emph{Through-Silicon-Vias (TSVs)}~\cite{jed:kee12}. As Figure~\ref{fig:hmc-struc} illustrates, the layers of the HMC are divided into 16 partitions, each of which is called a \emph{vault} with a corresponding memory controller in the logic layer, the so-called \emph{vault controller}~\cite{hybrid2013hybrid}. Each vault employs a 32-byte DRAM data bus~\cite{hybrid2013hybrid1}, enabled by 32 TSVs. A group of four vaults is called a \emph{quadrant}, connected to an external full-duplex serialized link, an eight- (half-width) or a 16-lane (full-width) connection clocking at speeds of 10, 12.5, or 15\,Gbps.  Thus, the maximum bandwidth of a two-link half-width HMC device with a 15\,Gbps link is:
\begin{equation}
\vspace{-1pt}
\label{eq:bw}
\small
  \text{BW}_\text{peak} = 2\,\text{link} \times
              8\,\nicefrac{\text{lanes}}{\text{link}} \times
             15\,\text{Gb/s} \times 2\,\small{\text{duplex}}
             = 60\,\text{GB/s.}
\end{equation}
\normalsize
\noindent The size of a DRAM layer in Gen2 (HMC\,1.1) devices is 4\,Gb. Since HMC\,1.1 has eight layers, the total size of it is 4\,GB. Moreover, each of the 16 vaults are 256\,MB. As the size of a bank is 16\,MB~\cite{hybrid2013hybrid1}, the number of banks in a vault and an HMC\,1.1 is 16 and 256, respectively.

\begin{figure}[t]
\centering
\vspace{-0.0in}
\includegraphics[width=0.85\linewidth]{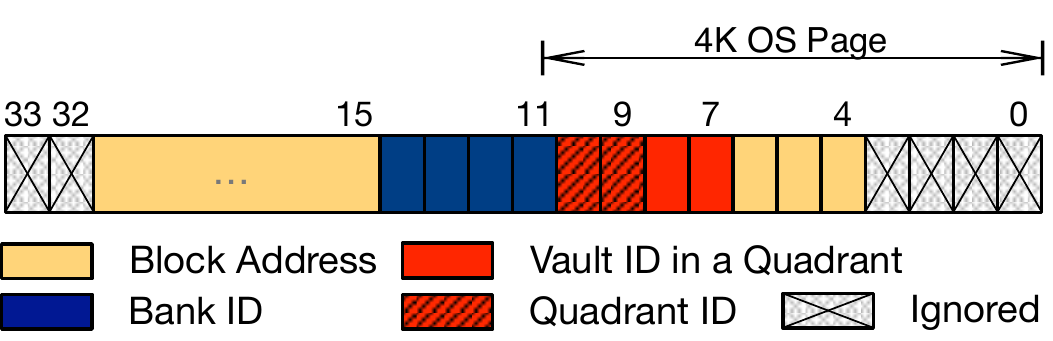}
\vspace{-0.0in}
\captionsetup{singlelinecheck=on,aboveskip=11pt,belowskip=-10pt}
\caption{Address mapping of 4\,GB HMC\,1.1 with block size of 128\,B).}
\label{fig:address-nocs}
\vspace{-0.0in}
\end{figure}

The header of an HMC\,1.1 request packet (see Section~\ref{sec:pck-mem} for more details) contains a 34-bit address field, but two high-order bits are ignored in a 4\,GB HMC. Figure~\ref{fig:address-nocs} shows the internal address mapping of HMC\,1.1 for 128\,B block size configuration~\cite{hybrid2013hybrid1}, as well as the \emph{low-order-interleaving} mapping of sequential blocks to vaults and then to banks within a vault. For a block size of 128\,B (Figure~\ref{fig:address-nocs}a), an OS page, usually 4\,KB, would be mapped to two banks over all 16 vaults so that serial accesses utilize bank-level parallelism (BLP). The vault controllers that each contain a part of a page are connected using an internal NoC, whose characteristics impacts the overall bandwidth and latency of the system.

%
\renewcommand{\arraystretch}{1}
\begin{table}[b]
    \vspace{-0.0in}
	\small
	\centering
	\vspace{-0.05in}
	\captionsetup{singlelinecheck=on,aboveskip=5pt}
	\caption{HMC request/response read/write sizes~\cite{hybrid2013hybrid1}.}
	\begin{tabular}{r| c c | c c}
		\toprule
		\multirow{2}{*}{\textbf{Type}} &
		\multicolumn{2}{c|}{\textbf{Request}} &
		\multicolumn{2}{c}{\textbf{Response}} \\
		& Read & Write & Read & Write \\
		\midrule
		Data Size & Empty & 1$\sim$8\,Flits & 1$\sim$8\,Flits & Empty\\
		Overhead & 1\,Flit & 1\,Flit & 1\,Flit & 1\,Flit   \\
		\midrule
		Total Size & 1\,Flit & 2$\sim$9\,Flits & 2$\sim$9\,Flits & 1\,Flit \\
		\bottomrule
	\end{tabular}
	\vspace{-0.0in}
	\vspace{0.0in}
	\label{tab:hmc-rdwrsize}
\end{table}
\renewcommand{\arraystretch}{1}
%

\subsection{Packet-Based Memories}
\label{sec:pck-mem}
Unlike memories with JEDEC-based bus interfaces (e.g., GDDR or HBM), HMC uses a packet-based interface to transfer packets over data links. Packet-based memories exploit internal and external NoCs for scalability; vaults in an HMC are connected internally and up to eight HMCs can be connected via external links. As the HMC interface uses high-speed serialization/deserialization (\emph{SerDes}) circuits, these networked implementations achieve higher raw link bandwidths than traditional, synchronous, bus-based interfaces. Unlike traditional memories, the access latency of a packet-based memory includes additional steps for packet processing, such as packet creation, port arbitration, flow control and serialization/deserialization~\cite{hybrid2013hybrid1}. These overheads are amortized by using large numbers of queues and ports (up to nine in our infrastructure) for sending packets, high BLP, and high-speed transmission to and from the HMC.

\begin{figure}[t]
\centering
\vspace{0.0in}
\includegraphics[width=1\linewidth]{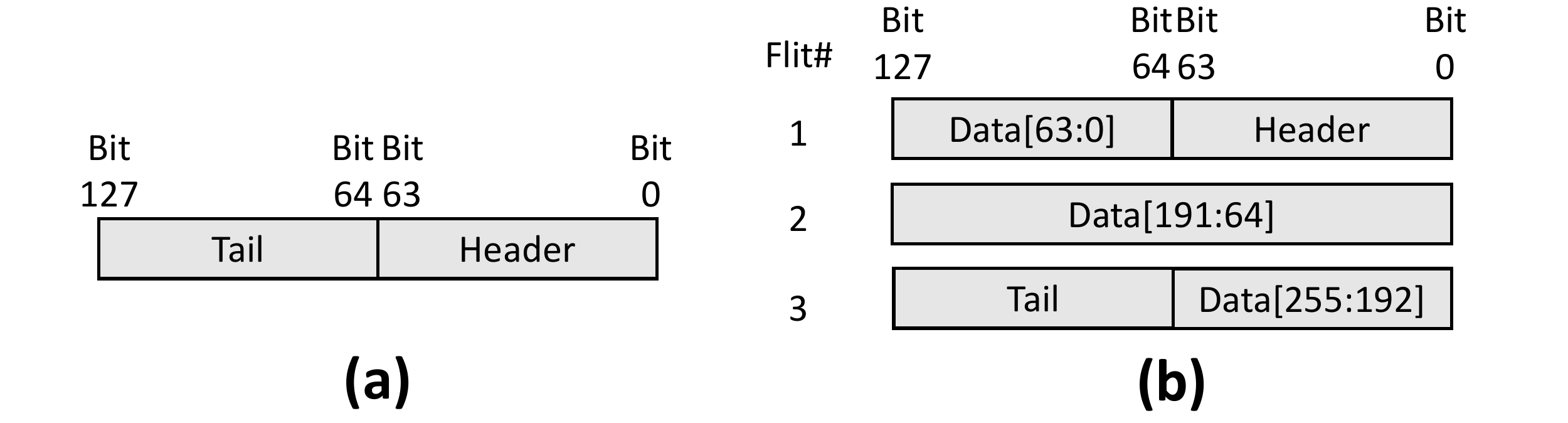}
\vspace{-0.0in}
\captionsetup{singlelinecheck=on,aboveskip=-3pt,belowskip=-7pt}
\caption{(a) A flow packet (no data), and (b) a request/response packet with 32B of data.}
\label{fig:hmc-pkt-format}
\vspace{-0.00in}
\end{figure}

Similar to IP-based networks, the communication of HMC is layered which includes physical, link, and transaction layers. The physical layer is responsible for serialization, deserialization, and transmission while the link layer handles low-level communication and flow control for packets over the high-speed physical connections. The transaction layer defines request and response packets, their fields, and controls high-level flow and retry. The HMC controller uses three types of packets: flow, request, and response packets. Flow packets do not contain a data payload while request and response packets are used for performing reads and writes from and to the HMC (Figure~\ref{fig:hmc-pkt-format}a and b). The 16-byte elements that construct packets are called \emph{flits}, and the size of data payload of each packet varies from one to eight flits. The least-significant flit of packets is transmitted first across the link. Flow control and integrity check of packets are performed using dedicated fields in the one-flit head and tail~\cite{hybrid2013hybrid1}. Accordingly, Table~\ref{tab:hmc-rdwrsize} shows each HMC transaction size in flits.

\begin{figure*}[]
  \vspace{-0.0in}
  \begin{tabular}{c | c}
  \begin{subfigure}{1\columnwidth}\centering
  \includegraphics[width=\textwidth]{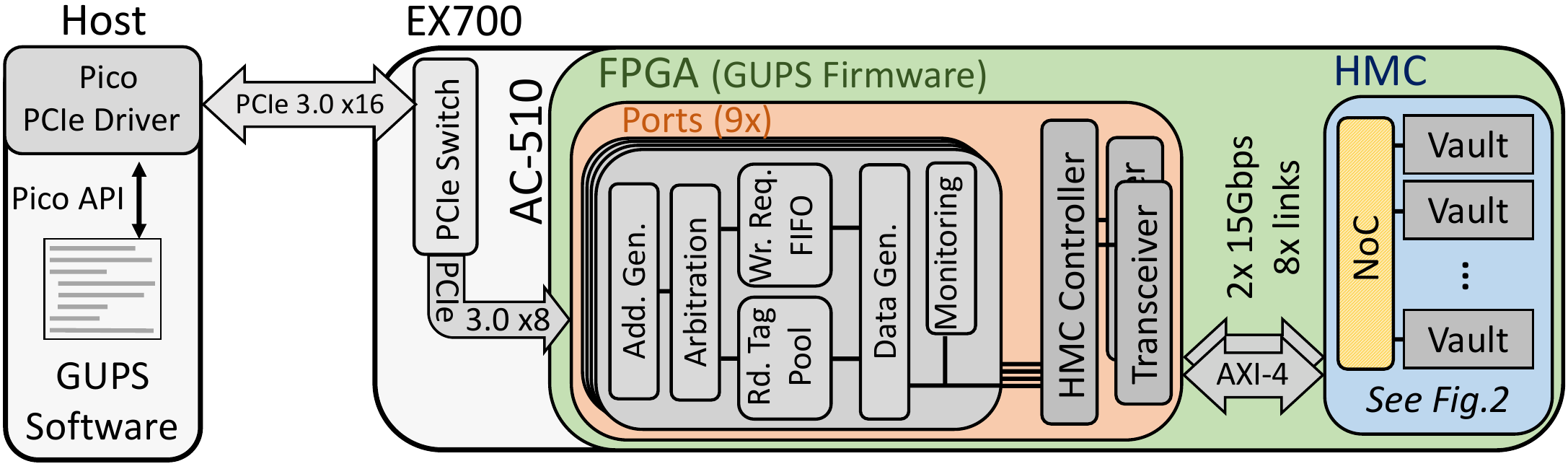}
  \captionsetup{singlelinecheck=on,aboveskip=8pt}
  \caption{}
  \label{fig:sys-gups}
  \end{subfigure}
  &
  \begin{subfigure}{1\columnwidth}\centering
  \includegraphics[width=\textwidth]{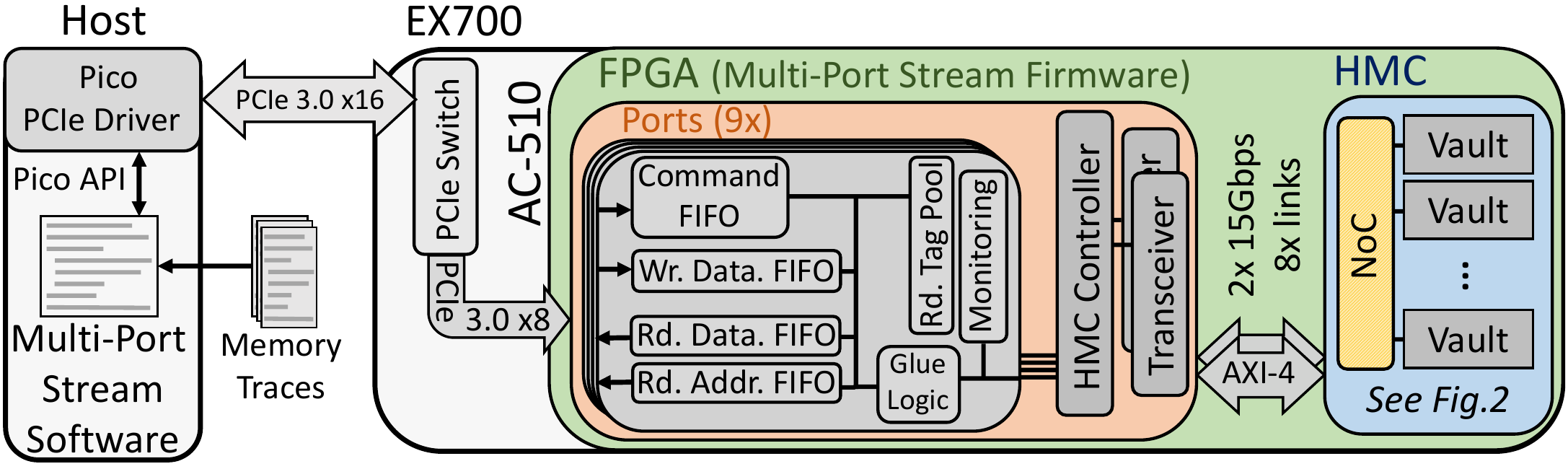}
  \captionsetup{singlelinecheck=on,aboveskip=8pt}
  \caption{}
  \label{fig:sys-multiport}
  \end{subfigure}
  \\
  \end{tabular}
  \captionsetup{singlelinecheck=on,aboveskip=4pt, belowskip=-5pt}
  \caption{Firmware and software overview: (a) GUPS implementation,
  and (b) multi-port stream implementation.}
  \vspace{-0.0in}
  \label{fig:sys-all}
\end{figure*}

\section{Methodology}
\label{sec:meth}

This section introduces our infrastructure for evaluating the HMC\,1.1 and includes details on its hardware, firmware (i.e., digital design on the FPGA), and software.

\subsection{Infrastructure}
We utilize a Pico SC-6 Mini~\cite{SC6Mini} machine that incorporates an EX-700~\cite{ex700} backplane, a PCIe\,3.0\,x16 board with 32\,GB/s bandwidth to the host. The EX-700 backplane can accommodate up to six AC-510~\cite{ac510} accelerator modules, each of which contains a Kintex Xilinx FPGA\footnotemark~and a 4\,GB HMC\,1.1 (similar to Figure~\ref{fig:hmc-struc}). We utilize one AC-510 in our evaluations. The HMC and the FPGA on an AC-510 module are connected with two half-width (8 lanes) links operating at 15\,Gbps, so the bi-directional peak bandwidth is 60\,GB/s, using Equation~\ref{eq:bw}.

\footnotetext{Part\#: xcku060-ffva1156-2-e}

\subsection{Firmware and Software}
We use two combinations of firmware and software to perform experiments, GUPS and multi-port stream implementations, shown in Figure~\ref{fig:sys-all}. Each combination integrates a custom logic on the FPGA, and a software counterpart. First, we describe the common components in the firmware on the FPGA. The FPGA uses Micron's HMC controller~\cite{hmc-controller} to generate packets for the multi-port AXI-4 interface between the FPGA and the HMC. On the software side, a Pico API and device driver are used to initialize the logic on the FPGA and provide an environment in which an OS communicates with the FPGA. The Pico API provides software methods to access the HMC through the FPGA with a direct path for sending/receiving packets. However, because the software runs at a lower rate on the host than on the FPGA, this solution cannot fully utilize the bandwidth of the HMC. Furthermore, since maximum frequency of the FPGA is low (187.5\,MHz), to generate more requests, the FPGA uses nine copy of the same module, called \emph{ports}. For measuring various statistics such as the total number of read and write requests and the total, minimum, and maximum of read latencies, each port contains a monitoring logic (not in the critical path of accesses). Note that monitoring logic measures aggregate latencies of the HMC controller, transceiver, data transmission on links, internal NoC, TSV transmission, and DRAM timings. Detailed studies of these latencies are performed in~\cite{had:asg17}, upon which we build our new measurements.

To observe the behavior of the NoC within the HMC with various traffic patterns and contention levels, we utilize two implementations as follows: (i) GUPS (Figure~\ref{fig:sys-gups}), a vendor-provided firmware that measures how frequently we can generate requests to random memory locations; and (ii) multi-port stream implementation (Figure~\ref{fig:sys-multiport}), a custom firmware which generates requests from memory trace files using Xilinx's AXI-Stream interface. 

{\em The GUPS implementation} is best suited to investigate NoC behavior under high contention while multi-port stream implementation does the same task from a trace file per port. For both implementations, the number of active ports and their access patterns are configured independently. With GUPS, each port has a configurable address generation unit that is able to send read-only, write-only, or read-modify-write requests for \emph{random}, or \emph{linear} mode of addressing. Also, by forcing some bits of the address to zero/one by using address \emph{mask}/\emph{anti-mask}, requests are mapped to a specific part of the HMC to create all possible access patterns (i.e., from accessing a single bank within a vault, to accessing all banks of all the vaults). To perform experiments, for each port, we first set the type of requests and size, their mask and anti-mask, and then we activate the port. While the port is active, it generates as many requests as possible for 10 seconds, and then it reports the total number of accesses (read and write), maximum/minimum of read latencies, and aggregate read latency back to the host. In this paper, the type of requests are read only, unless stated otherwise. Our current firmware implementations do not support ACKs after writes, so accurate measurements of write latency would only be possible with added monitoring logic specifically for writes. We plan to address this limitation in future work. However, since we study the internal NoC of the HMC, any type of requests that consume resources will reveal the behavior, bottleneck, and impacts of the NoC.

{\em The multi-port stream implementation} employs a multi-threaded software that reads a memory trace file for each port and populates buffers on the host. Then, by using Xilinx's AXI-Stream interface to each port (wrapped in a PicoStream API call \cite{pico-api}), we efficiently transmit commands such as access types, sizes, and data through their dedicated communication channel. After issuing requests and waiting for responses, each port transmits read data and their addresses back to the host. In fact, the FPGA reads sent data continuously until it reads all of it, such that each port reads data from its dedicated channel in every cycle. In both GUPS and multi-port stream implementations, we calculate the average access latency of reads by dividing the aggregate read latency by the total number of reads. We calculate bandwidth by multiplying the number of accesses by the cumulative size of request and response packets including header, tail and data payload (shown in Table~\ref{tab:hmc-rdwrsize}), and dividing it by the elapsed time.

\section{Results}
\label{sec:res}

This section presents various detailed latency and bandwidth analyses under various traffic conditions and access patterns with GUPS and multi-port stream implementations.

\subsection{High-Contention Access Latency}
\label{sec:res-high}

\begin{figure}[b]
\centering
\vspace{-4pt}
\includegraphics[width=1\linewidth]{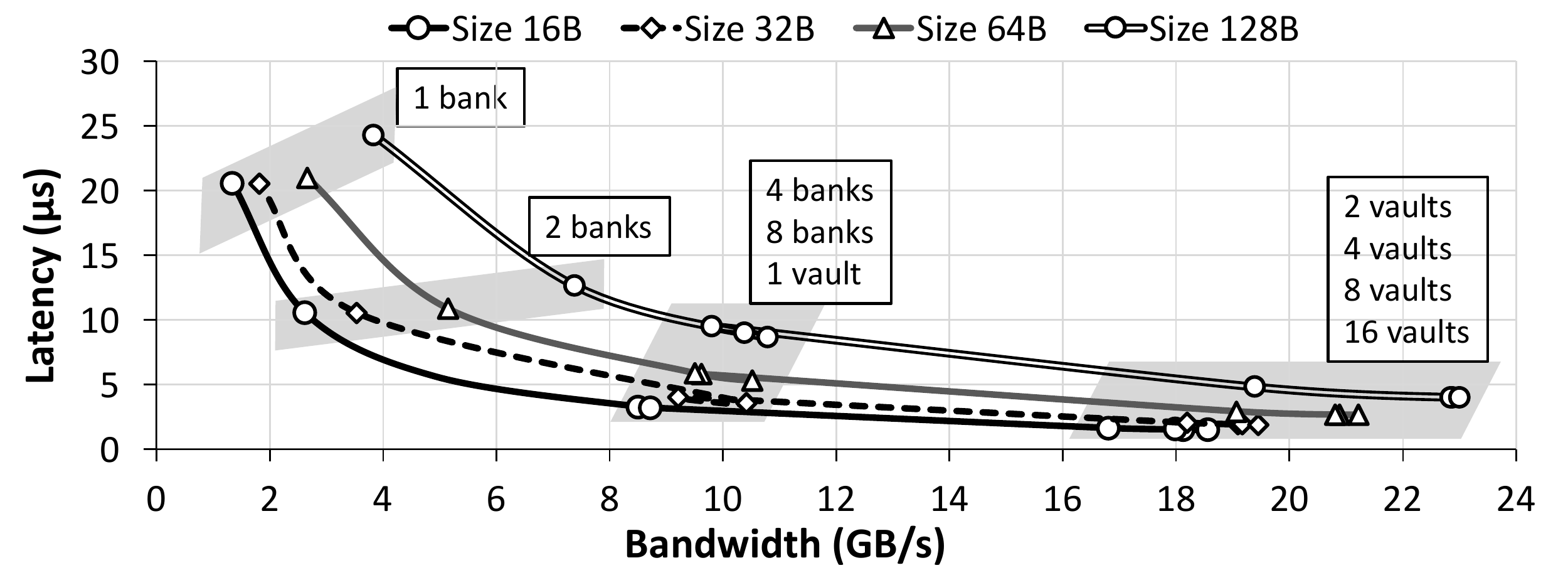}
\captionsetup{singlelinecheck=on,aboveskip=5pt,belowskip=0pt}
\caption{Read latency and bi-directional bandwidth for various access patterns and request sizes for read-only requests. }
\label{fig:bw-lat}
\vspace{-0.0in}
\end{figure}

To achieve a broad perspective of the HMC properties, we perform experiments that access various structural organizations in the HMC, such as vaults and banks. Figure~\ref{fig:bw-lat} illustrates the latency and bandwidth relationship for read-only accesses. The lowest bandwidth for undistributed accesses (i.e., accesses targeted to a bank) is 2\,GB/s for 32\,B requests, and the highest bandwidth for the most distributed accesses (i.e., accessing more than or equal to two vaults) is 23\,GB/s for 128\,B requests. Note that, as Table~\ref{tab:hmc-rdwrsize} shows, read-only requests mostly utilize response bandwidth, which has a cap of 30\,GB/s. Accesses to more than two vaults has a constant bandwidth, caused by the limitation of the external link bandwidth between the HMC and the FPGA. Moreover, accesses distributed over eight banks, but within one vault, have the same 10\,GB/s bandwidth, limited by the maximum internal bandwidth of a vault~\cite{ros14}. Figure~\ref{fig:bw-lat} also shows that as the accesses become less distributed, the latency of accesses increases. As the figure depicts, access latency varies from 24,233\,ns for 128\,B requests targeting a single bank, to 1,966\,ns for 16\,B requests spread across more than two vaults. Less distributed access patterns (e.g., one bank) have higher latency because they benefit less from BLP. Furthermore, the latency of small requests is always lower than that of large requests because (\romannum{1}) the granularity of the DRAM bus within each vault is 32\,B~\cite{hybrid2013hybrid1}, so data payloads larger than 32\,B is split; and (\romannum{2}) larger request packets constitute more flits, so buffering and reordering of packets cause higher latencies.

Figure~\ref{fig:bw-lat} illustrates that large requests (e.g., 128\,B) always have higher bandwidth utilization than that of small requests (e.g., 32\,B). This is because (\romannum{1}) large packets utilize bandwidth more effectively (i.e., less overhead), and (\romannum{2}) small requests quickly consume the maximum number of tags of outstanding requests to the HMC. As Table~\ref{tab:hmc-rdwrsize} presents, each packet, regardless of its data size, always has an overhead of one flit (i.e., 16\,B). For this reason, the bandwidth efficiency of read responses with 16\,B and 128\,B data sizes are $\nicefrac{16}{16+16}=50\%$ and $\nicefrac{128}{128+16}=89\%$, respectively. Moreover, for retransmission of a packet (because of transmission failure, flow control, or CRC failure), each port must track outstanding requests, so each port can handle a limited number of outstanding requests at a time. Small requests, compared to large requests, underutilize this limited number of slots for keeping smaller data, which results in low bandwidth utilization.
In summary, large packet sizes utilize available bandwidth more effectively at the cost of added latency. In addition, for reducing access latency, accesses should be carefully distributed to exploit BLP and avoid bottlenecks.

\subsection{Low-Contention Access Latency}
\label{sec:res-low}

\begin{figure}[t]
\centering
\vspace{-0.0in}
\includegraphics[width=1\linewidth]{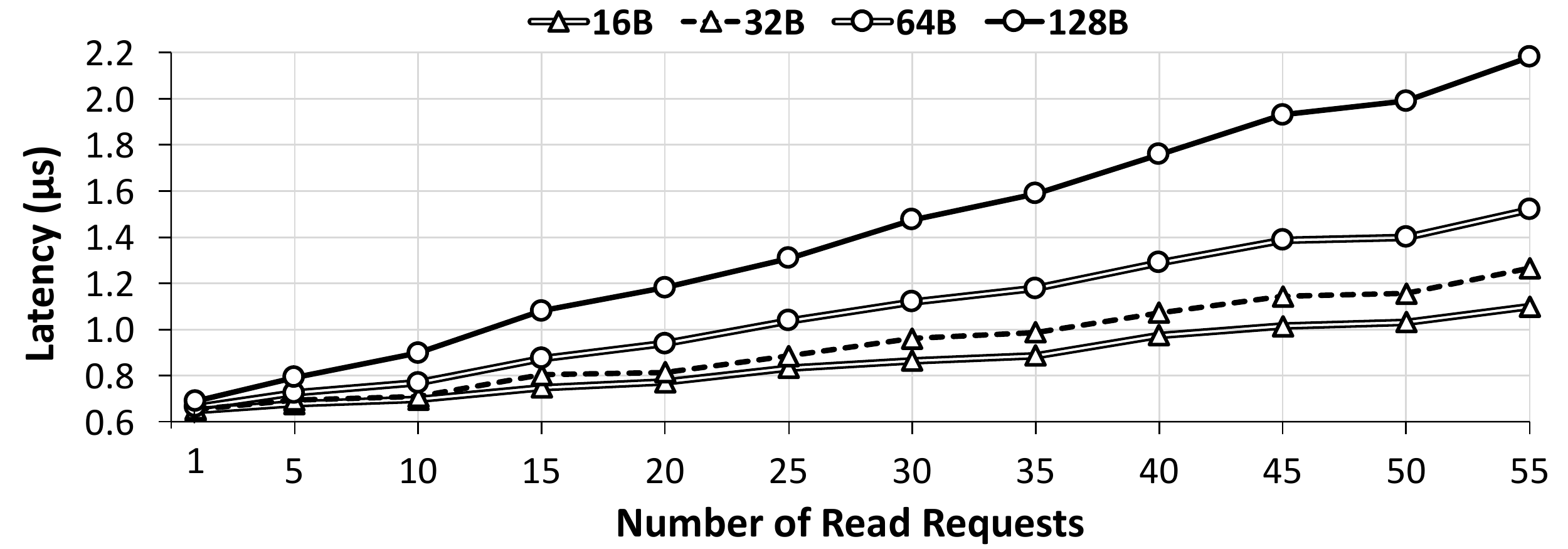}
\vspace{-0.0in}
\captionsetup{singlelinecheck=on,aboveskip=-6pt,belowskip=-5pt}
\caption{Average latency of low-load accesses for various request sizes for the number of requests in the range of one to 55.}
\label{fig:low-load-lat}
\vspace{-0.1in}
\end{figure}

To examine low-contention latencies, we measure the access latency of the HMC while limiting the number of random read requests to be mapped within the 16 banks of a vault. Then, for each number of read requests, we report average latency across all vaults. To tune the number of accesses and the size of request packets, we use the multi-port stream implementation. Figure~\ref{fig:low-load-lat} depicts that as the number of requests in a stream (stream in this context means a limited number of requests) increases from one to 55, the average latency increases from 0.7 to 1.1\,$\mu$s for the request size of 16\,B, and from 0.7 to 2.2\,$\mu$s for the request size of 128\,B. In other words, we observe two behaviors: (\romannum{1}) when the number of request packets is small, the size of request packet does not effect the latency; and (\romannum{2}) when the size of request packets is larger, the requests experiences more variations in the latency. Since the flow control unit in the infrastructure is only activated with a large number of outstanding requests, we are certain that, as reported in~\cite{had:asg17}, approximately 547\,ns of all latencies for the small number of requests in Figure~\ref{fig:low-load-lat} belongs to FPGA and data transmission stages. Therefore, the contributing latency of HMC under low load (i.e., no load) is 100 to 180\,ns, which includes the latency of DRAM timings ($t_{RCD}+t_{CL}+t_{RP}$ is around 41\,ns for HMC~\cite{kim:kim13,ros14}), TSV transmission, vault controller, and internal NoC. However, as the number of requests increases, with the same BLP, queuing delay in both the HMC (i.e., internal NoC and vault controllers) and the FPGA increases, which results in an order of magnitude higher delays. Note that since HMC utilizes a packet-switched interface to vault controllers in its logic layer, the observed average latency of the HMC is higher than that of traditional DDRx.

\begin{figure}[b]
\centering
\vspace{-5pt}
\includegraphics[width=1\linewidth]{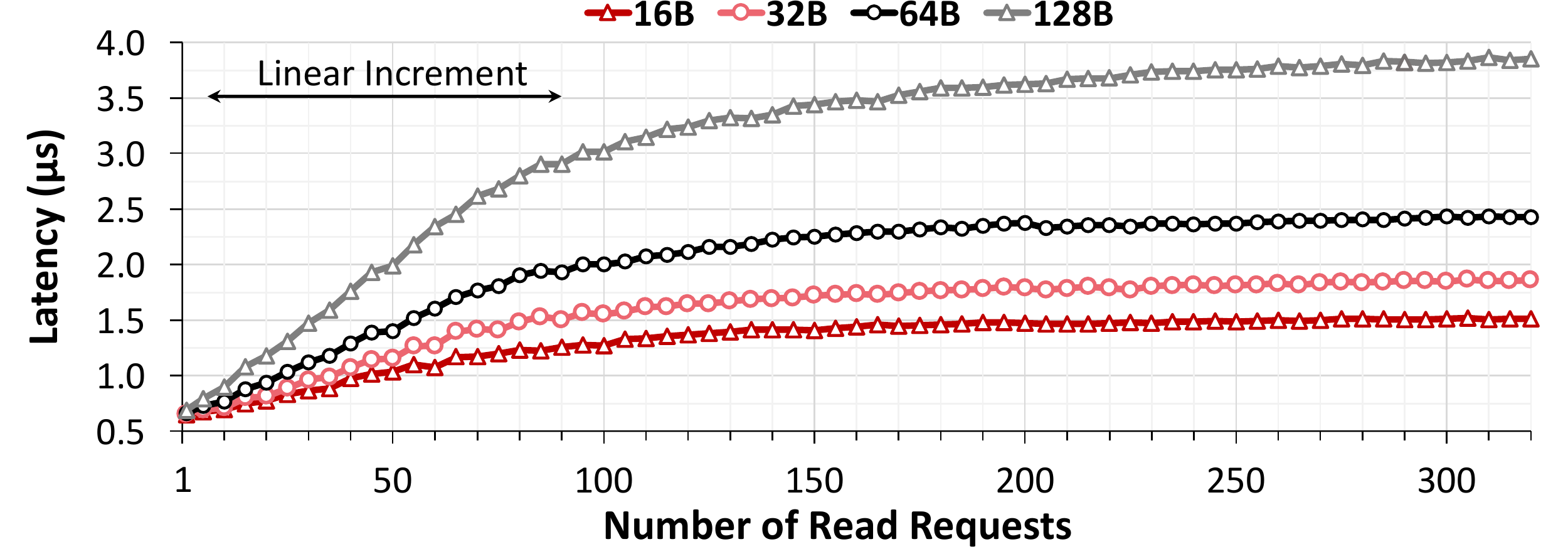}
\vspace{-0.0in}
\captionsetup{singlelinecheck=on,aboveskip=-4pt,belowskip=0pt}
\caption{Average latency of low-load accesses
for various request sizes for the number of requests in the range of one to 350.}
\label{fig:low-load-lat-broad}
\vspace{-0.0in}
\end{figure}

Figure~\ref{fig:low-load-lat-broad} illustrates a wider range for the number of read requests in a stream than that shown in Figure~\ref{fig:low-load-lat}. In this figure, we observe that when the number of requests increases up to 100, average access latency increases linearly. After that, the latency stays approximately constant when the number of requests grows. By assuming a hypothetical queue for requests, we infer that until the time that the queue is not full, the latency of each request equals to its serving time plus its waiting time, which is the sum of the serving time of all previous requests that are already in the queue. We can write the average latency of $n$ requests as $\nicefrac{\sum_{i=0}^{n}(iS)}{n}$, in which $S$ is the serving time of a request. Therefore, the latency seen by each request is correlated to the number of requests in the queue. In the region, where latency remains constant, the queue is always full, so the latency of a request equals to it serving time plus the waiting time for all requests in a the queue (i.e., $n=\text{Queue}_{\text{Size}}$). Thus, the linear region represents a partially utilized system, and the constant region represents a fully utilized system. Section~\ref{sec:res-req-res-bw} provides further details on bandwidth and bottlenecks. From the system perspective, the linear region achieves a lower latency while providing less bandwidth than that of the saturated region. Thus, based on the sensitivity of an application to the latency, a system may exploit these two regions to gain performance. To recap, even for low-contention traffics, NoC and queuing delay contribute significantly to the access latency of the HMC, and therefore broadly to the performance.

\begin{figure}[b]
  \centering
  \vspace{-0.1in}
  \begin{subfigure}{0.98\columnwidth}
  \includegraphics[width=\textwidth]{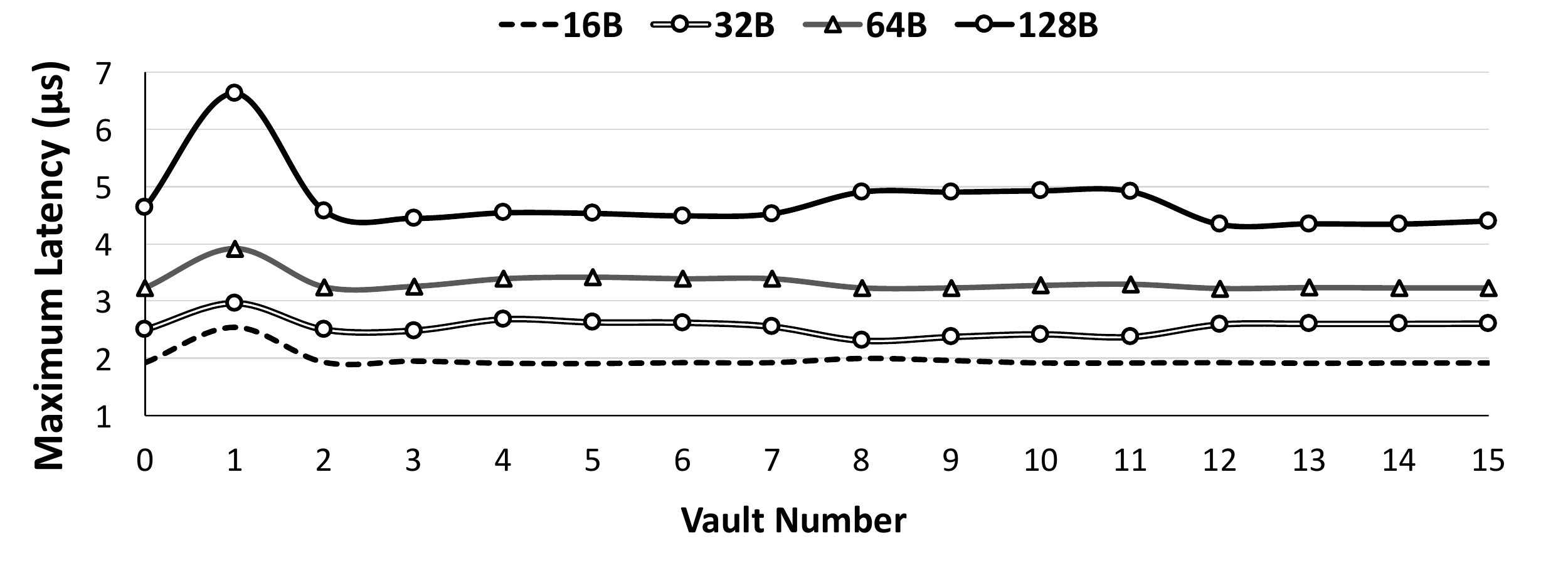}
  \captionsetup{singlelinecheck=on,aboveskip=-2pt,belowskip=0pt}
  \caption{}
  \end{subfigure}
  
  \begin{subfigure}{0.98\columnwidth}
  \includegraphics[width=\textwidth]{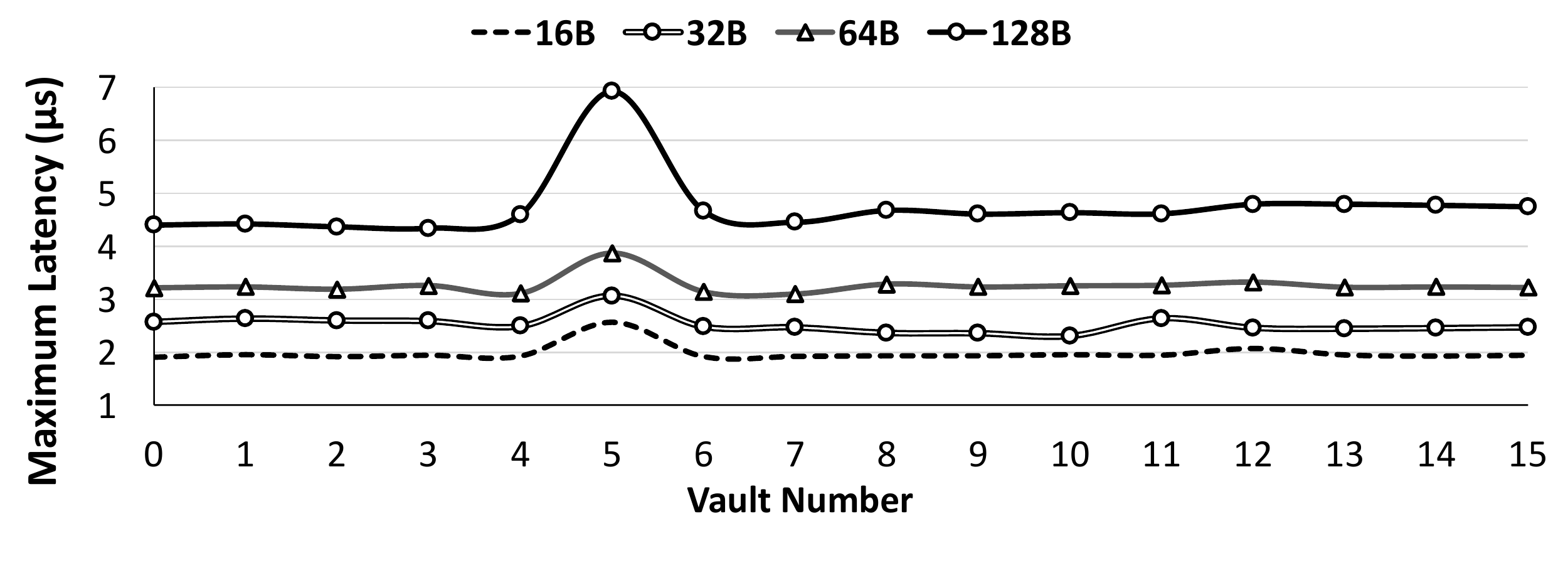}
  \captionsetup{singlelinecheck=on,aboveskip=-4pt,belowskip=0pt}
  \caption{}
  \end{subfigure}
  
  \captionsetup{singlelinecheck=on,aboveskip=2pt,belowskip=0pt}
  \caption{Maximum observed latency in accessing four vaults, three of which are the same. Accessing vault numbers (a) one (3x) and all vaults; and (b) five (3x) and all vaults.}
  \label{fig:qos-raw}
\end{figure}

\begin{figure*}[]
  \vspace{0.0in}
  \begin{tabular}{l l l l}
  \begin{subfigure}{0.48\columnwidth}\centering
  \includegraphics[width=\textwidth]{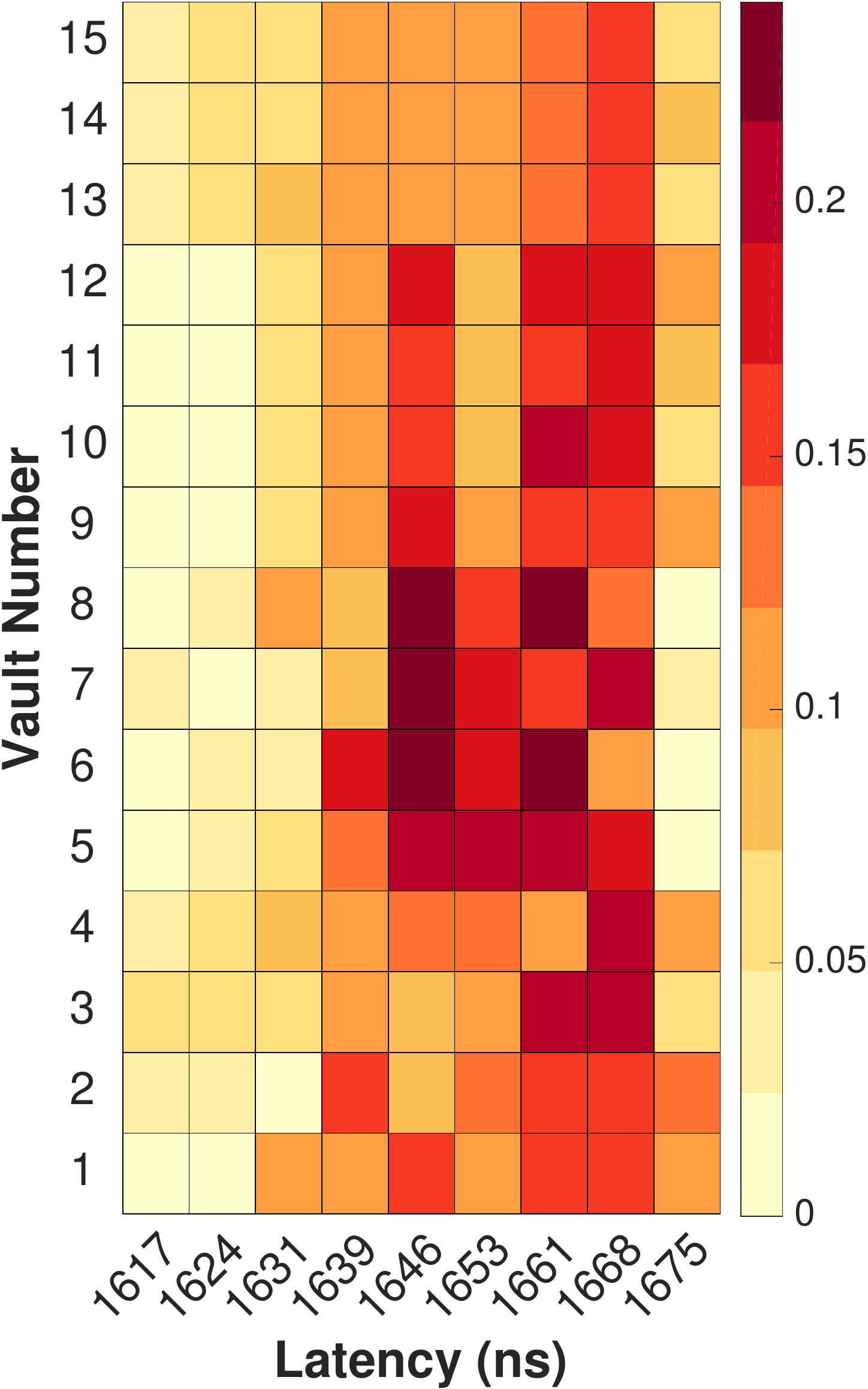}
  \captionsetup{singlelinecheck=on,aboveskip=4pt,belowskip=3pt}
  \caption{16B}
  \label{}
  \end{subfigure}
  &
  \begin{subfigure}{0.48\columnwidth}\centering
  \includegraphics[width=\textwidth]{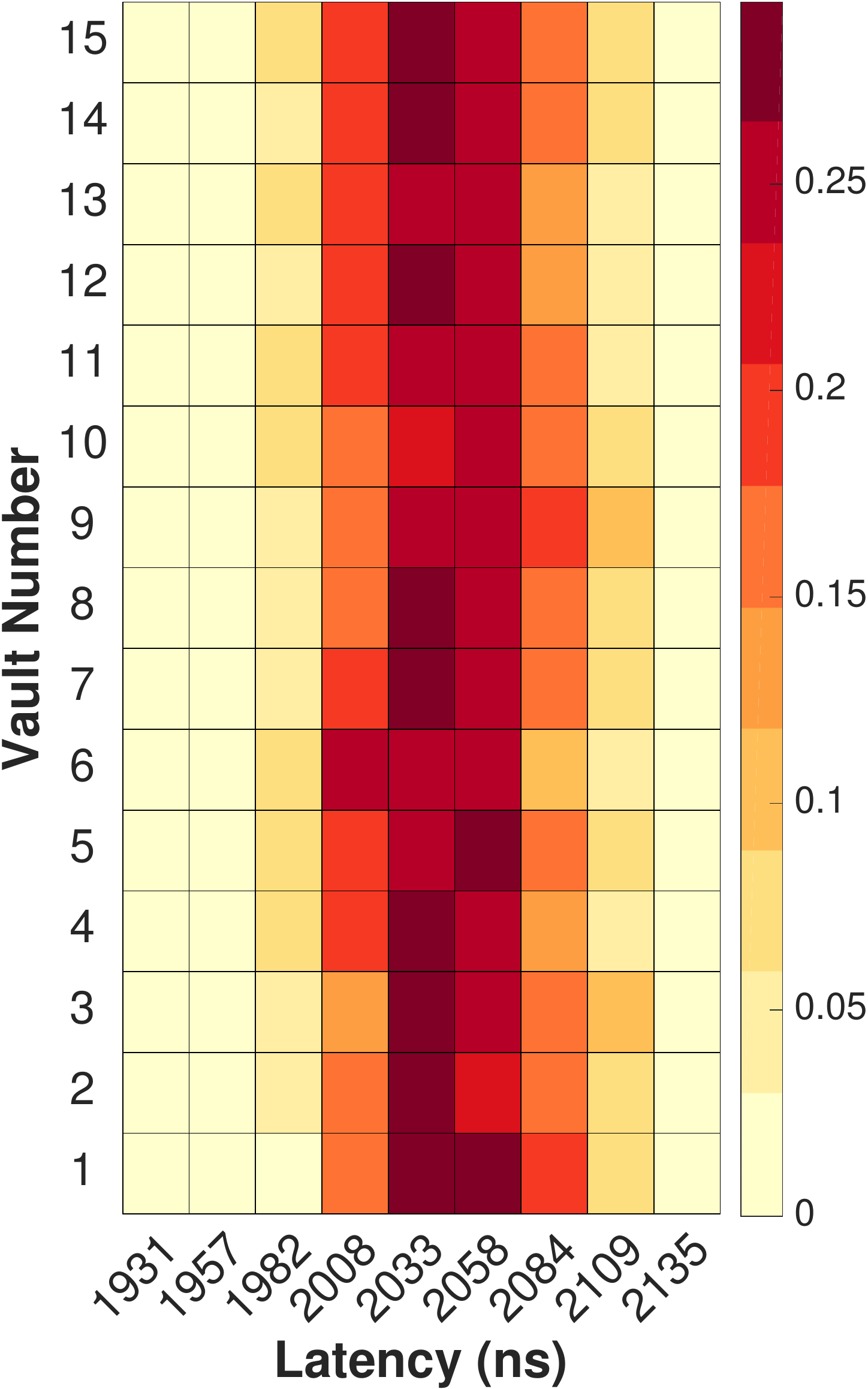}
  \captionsetup{singlelinecheck=on,aboveskip=4pt,belowskip=3pt}
  \caption{32B}
  \label{}
  \end{subfigure}
  &
  \begin{subfigure}{0.48\columnwidth}\centering
  \includegraphics[width=\textwidth]{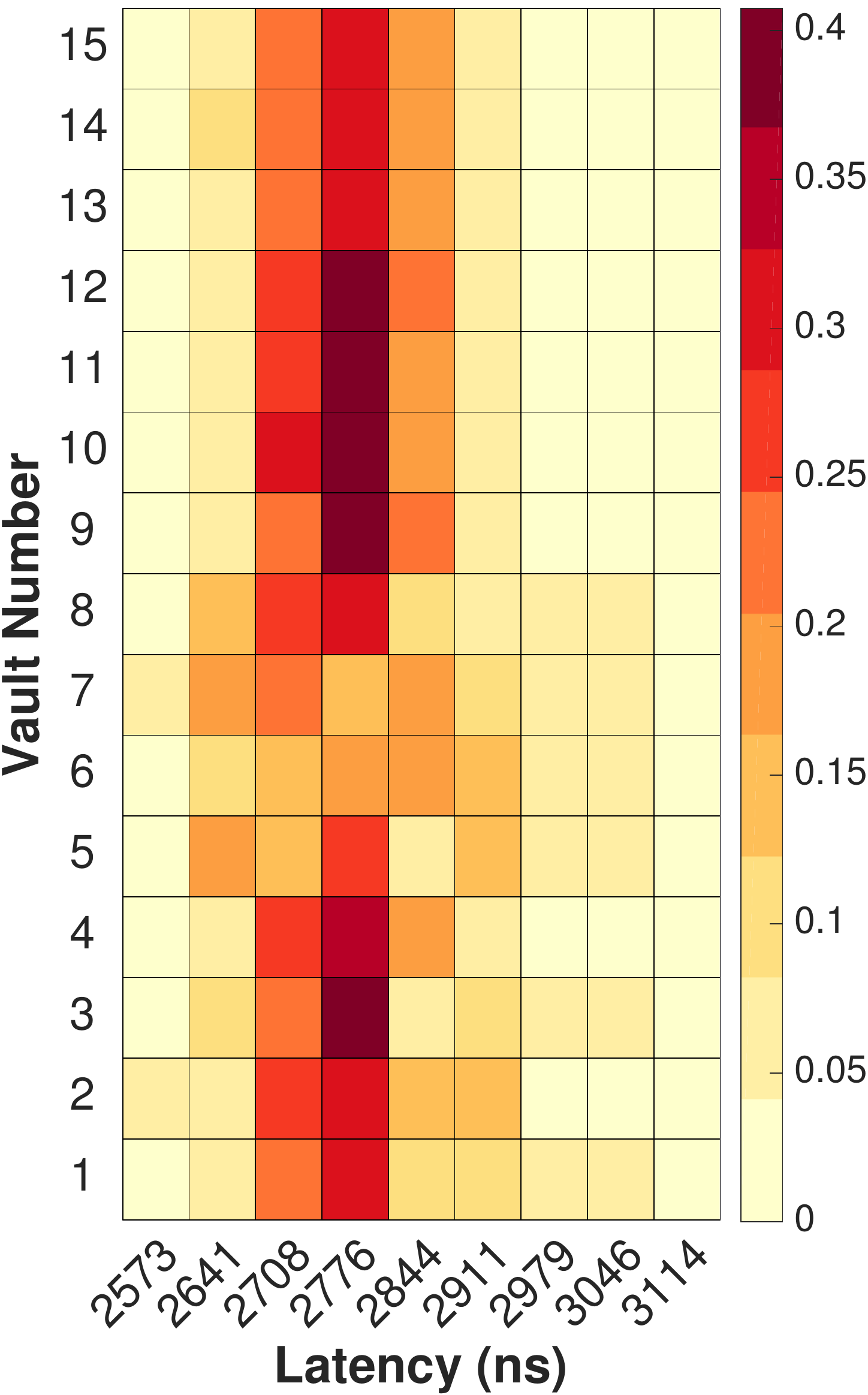}
  \captionsetup{singlelinecheck=on,aboveskip=4pt,belowskip=3pt}
  \caption{64B}
  \label{}
  \end{subfigure}
  &
  \begin{subfigure}{0.48\columnwidth}\centering
  \includegraphics[width=\textwidth]{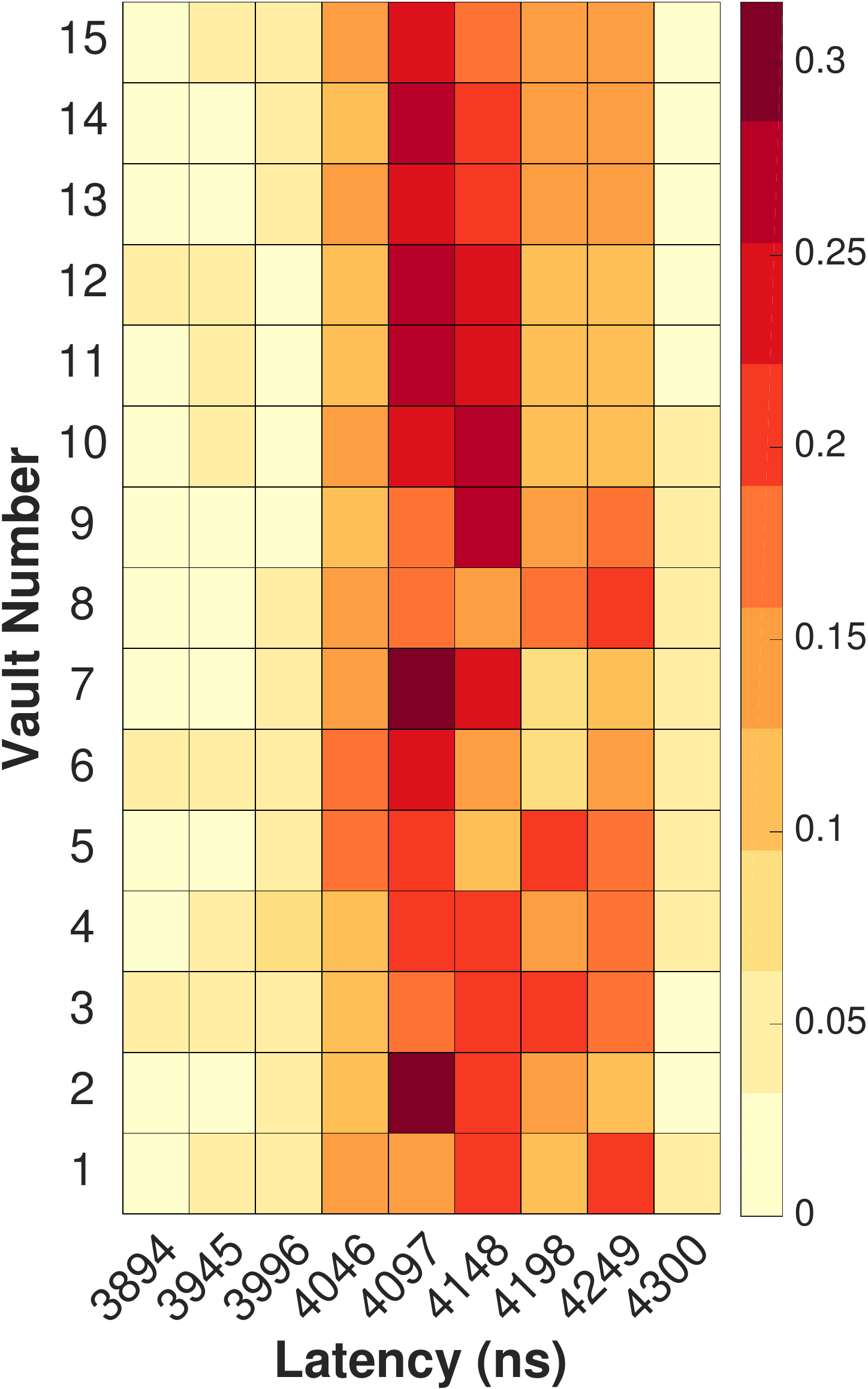}
  \captionsetup{singlelinecheck=on,aboveskip=4pt,belowskip=3pt}
  \caption{128B}
  \label{}
  \end{subfigure}
  \\
  \end{tabular}
  \captionsetup{singlelinecheck=on,aboveskip=3pt,belowskip=-8pt}
  \caption{Latency histograms of each vault in heatmaps for various request sizes of (a) 16, (b) 32, (c) 64, and (d) 128\,B.}
  \vspace{-0.0in}
  \label{fig:histgram-vault}
\end{figure*}

\subsection{Quality of Service Analysis}
\label{sec:res-qos}

Similar to other networks, QoS of a packet-switched-based memory refers to guaranteeing the required latency or bandwidth for an application. In this section, we inspect techniques to manage the resources in a packet-switched memory to achieve required QoS.  In particular, our goal is to ascertain how latency varies within an access pattern (e.g, accesses distributed in four vaults) as a result of the packet-switched interface of the HMC, and subsequently, how this will affect the QoS of applications. The effects of latency variations on QoS are important because they impact latency-sensitive applications~\cite{iye:zha07}, QoS guarantees~\cite{dean:bar13}, denial of service~\cite{mos:mut07}, and multi-threaded and parallel architectures that stall for the slowest thread (i.e., work imbalance). A packet-switched memory, despite its high bandwidth (thanks in part to serialization, and high BLP in a small area), adds uncertainty to access latencies. Therefore, as we will see, only optimizing the accessing patterns to the HMC in an application would not be sufficient to guarantee a precise QoS.

In our experiments, as a case study, we use four ports with the multi-port stream implementation to generate read accesses to four vaults (targeting 1\,GB in total). During the experiments, three ports always access the same vaults, and the fourth port iterates over all possible vaults. Figure~\ref{fig:qos-raw}a and b illustrates the maximum observed latency for two series of experiments, in which the three ports always access vault number one and five, respectively. The figures depict when the accesses of the forth port are to the same vault as the other ports (i.e., vault numbers one and five in Figures~\ref{fig:qos-raw}a and b, respectively) the maximum observed latency increases up to 40\% relative to other accesses. Furthermore, when the forth port is not accessing the same vault, maximum observed latency varies notably. For instance, the maximum variations are around 200, 330, 400, 600\,ns for 16, 32, 64, and 128\,B size of requests, respectively. 

In summary, even within the same access pattern, NoC causes considerable latency variations, which will have a noticeable impact on QoS of an application, even when its access patterns are optimized. Note that Figure~\ref{fig:qos-raw} illustrates results for only four ports, and if number of ports (i.e., threads or applications) accessing one vault increase, the latency would increase even more. This general trend in latency increment helps providing an approximate QoS for various traffic conditions with diverse latency requirement. For instance, in a case that we have five traffic streams, four of which can be served in long latency, and one has high priority and requires a fast service; the system can assign a limited number of vaults to all four low-priority traffic streams, and remaining vaults to the high-priority traffic. Therefore, the QoS of all traffic streams would be satisfied. Such techniques for managing QoS can be provided in the host-side memory controller by real-time remapping, or reserving resources.

\subsection{High-Contention Latency Histograms Per Vault}
\label{sec:res-hist-latency}

To understand the impact of accessing various combinations of vaults on performance, we extend the experiments of the previous section, which accessed four vaults using the multi-port stream implementation. For instance, accesses to four consecutive vaults (e.g., 0,1,2, and 3) that share network resources may have higher latency than accesses spread among non-consecutive vaults (e.g., 0,4,8, and 12) do. To test this hypothesis, we access all possible combinations of four different vaults (i.e., equal to 1820 combinations, or $\nicefrac{n!}{k! \times (n-k)!}$ for $n=16$ and $k=4$) with various request sizes and calculate the average access latency among four vaults. Then, we associate the calculated average latency with every vault in that combination. 

\begin{figure}[b]
\centering
\vspace{-8pt}
\includegraphics[width=1\linewidth]{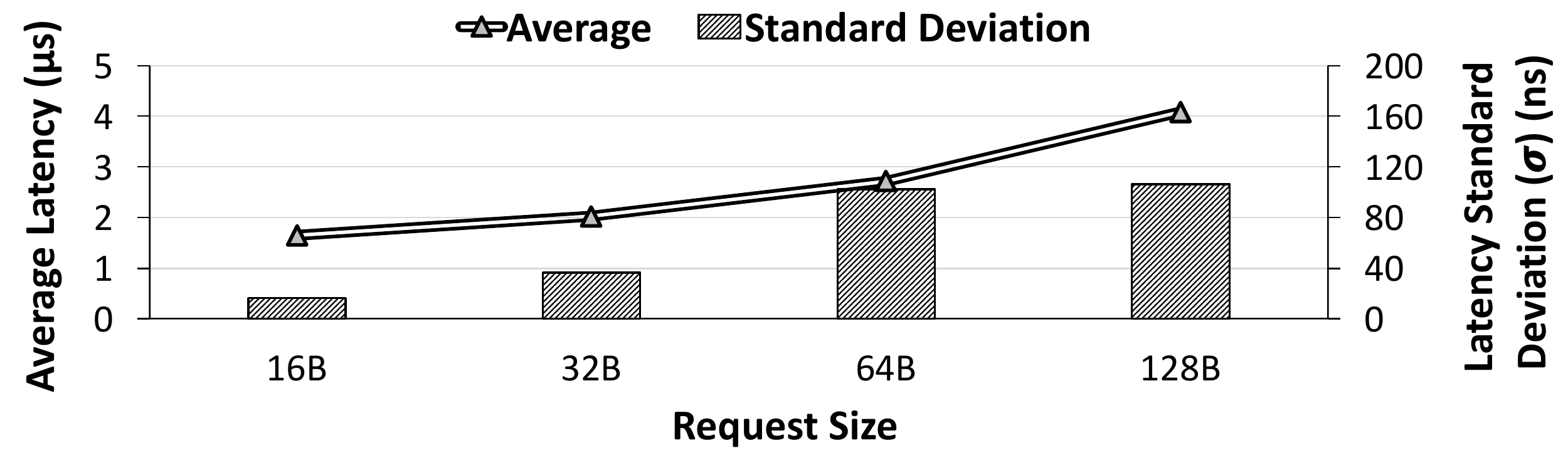}
\vspace{-0.0in}
\captionsetup{singlelinecheck=on,aboveskip=-8pt,belowskip=0pt}
\caption{The average and standard deviation of latency across all vaults for various sizes in the four-vault access pattern.}
\label{fig:hist-latency-avg}
\vspace{-0.0in}
\end{figure}

\begin{figure*}[h]
  \vspace{-0.0in}
  \begin{tabular}{l l}
  \begin{subfigure}{0.98\columnwidth}\centering
  \includegraphics[width=\textwidth]{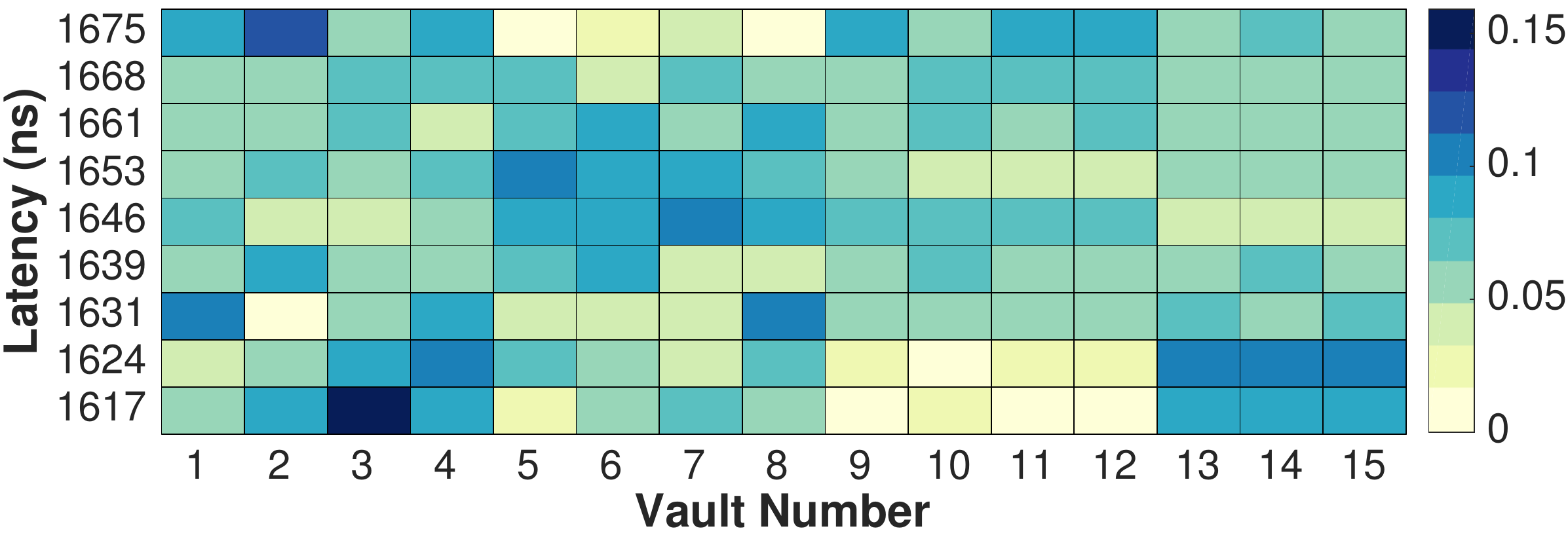}
  \captionsetup{singlelinecheck=on,aboveskip=0pt,belowskip=8pt}
  \caption{16B}
  \end{subfigure}
  &
  \begin{subfigure}{0.98\columnwidth}\centering
  \includegraphics[width=\textwidth]{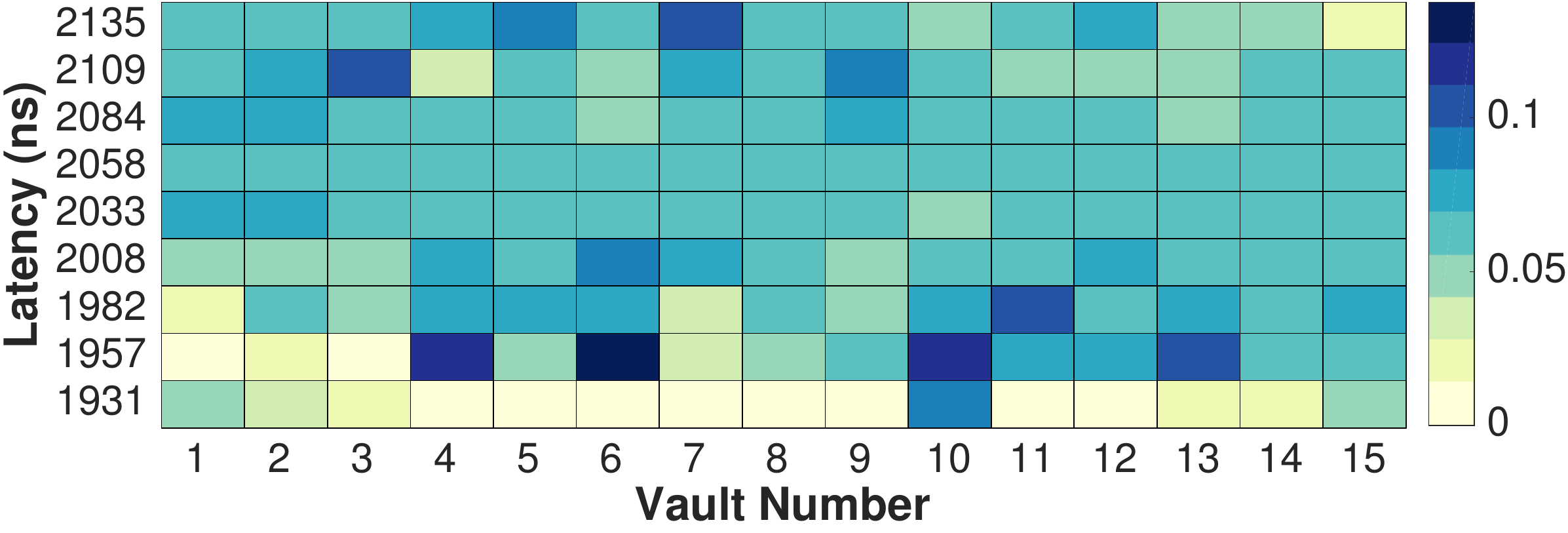}
  \captionsetup{singlelinecheck=on,aboveskip=0pt,belowskip=8pt}
  \caption{32B}
  \end{subfigure}
  \\
  \begin{subfigure}{0.98\columnwidth}\centering
  \includegraphics[width=\textwidth]{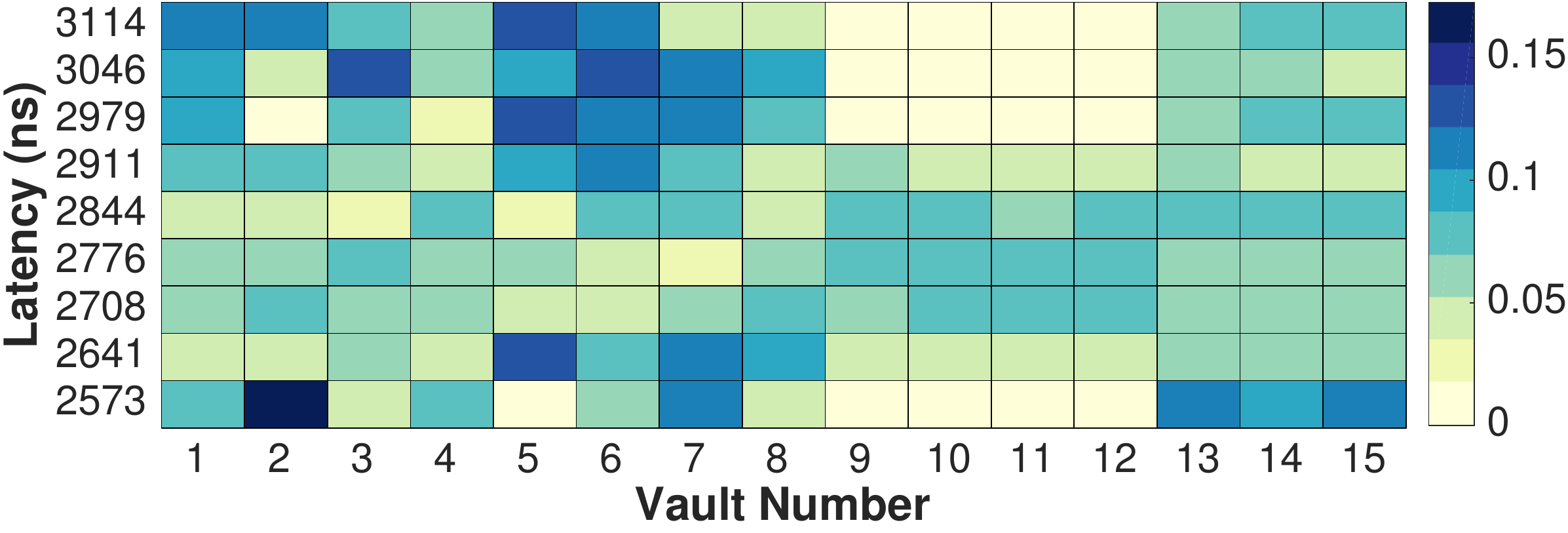}
  \captionsetup{singlelinecheck=on,aboveskip=0pt,belowskip=8pt}
  \caption{64B}
  \end{subfigure}
  &
  \begin{subfigure}{0.98\columnwidth}\centering
  \includegraphics[width=\textwidth]{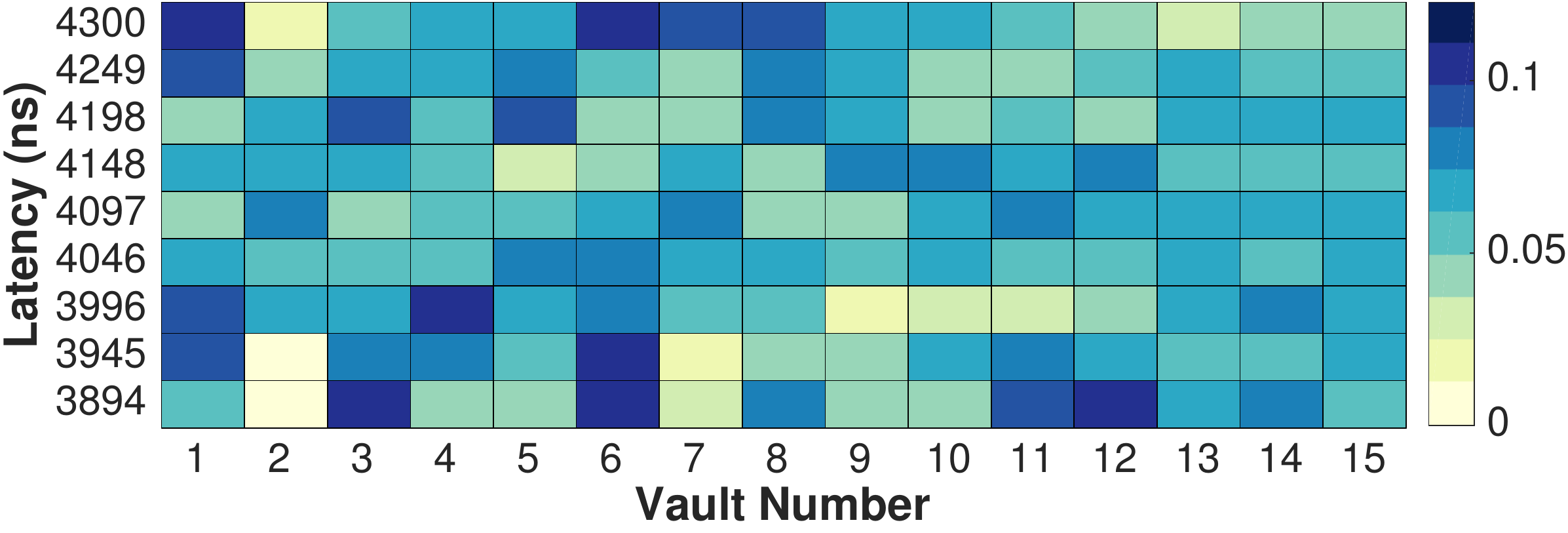}
  \captionsetup{singlelinecheck=on,aboveskip=0pt,belowskip=8pt}
  \caption{128B}
  \end{subfigure}
  \\
  \end{tabular}
  \captionsetup{singlelinecheck=on,aboveskip=-0pt,belowskip=-10pt}
  \caption{Vault histograms of each latency interval in heatmaps for various  request sizes of (a) 16, (b) 32, (c) 64, and (d) 128\,B.}
  \vspace{-0.0in}
  \label{fig:histgram-latency}
\end{figure*}

Figure~\ref{fig:histgram-vault} illustrates our results for various sizes in heatmaps where a row represent the latency histogram of a vault. In other words, in a row, the color of a rectangle represents the normalized value of the number of accesses in that latency interval against the total number of accesses to the corresponding vault (i.e., 455). As the figure shows, each vault has a different latency behavior. For instance, in Figure~\ref{fig:histgram-vault}c, we observe that the histogram of vaults differs substantially (e.g., vault numbers 5, 6, and 7). Although we can investigate these figures in more detail, a quick takeaway is that purely optimizing the general access patterns (in our example, four-vault access pattern) of an application would not guarantee a particular latency. In other words, Figure~\ref{fig:histgram-vault} presents a case study with a four vault access pattern, in which the only factor of variation is the number of the vault that determine the physical location of a vault within the 3D stack. Therefore, since other factors such as access pattern are same, we conclude that the NoC design of the HMC has a significant impact on the observed latency variations.
    
As Figure~\ref{fig:histgram-vault} shows, for each request size, although all the vaults have a similar average latency, the distribution of latencies are different among vaults. For a better illustration, Figure~\ref{fig:hist-latency-avg} depicts the average latency of all vaults and the standard deviation for various packet sizes. We observe that the standard deviation of latencies is 20, 40, 100, and 106\,ns for request sizes of 16, 32, 64, and 128\,B, respectively. Note that 68\% of a population is within $(\mu+\sigma,\,\mu-\sigma)$, in which $\mu$ and $\sigma$ are average and the standard deviation of that population, respectively. For a particular request size, while the average latency per vault are similar, the distribution of it per vaults covers a broad range. Compared to smaller request sizes, larger request sizes have more variations in latency, because large request sizes occupy larger buffer spaces than small request sizes do. Also, large requests incur extra delays because of reordering and packetizing. Therefore, small request sizes are good candidates for guaranteeing a high QoS. However, as discussed in Section~\ref{sec:res-high}, small request sizes have low bandwidth efficiency and generally provide lower bandwidth utilization than large request sizes.

In detail, we infer the following insights from Figure~\ref{fig:histgram-vault}: (\romannum{1}) Comparing the four subfigures, which indicate the latency for various packet sizes, shows that when the size of requests increases, the latency increases. For instance, the latency of 128\,B accesses is in range of 4\,$\mu$s, which is 2.5x as long as that of 16\,B accesses. A recent paper~\cite{had:asg17} observes a similar behavior in a limited experiment in accessing to a random vault and conclude such variations is caused by the granularity of 32\,B DRAM bus within a vault. (\romannum{2}) The range of the latency variations for 16, 32, 64, and 128\,B accesses are 29, 76, 136, and 203\,ns, which indicates that the smallest requests have more consistent latency, and largest requests have more variable latency. Therefore, the size of accesses also impacts the variation in latency. (\romannum{3}) By comparing the latency of each vault from the rows of each subfigure, we see that each vault have a random behavior and we cannot allocate a specific latency to a vault based on its location (i.e., number). In other words, the latency of each vault is impacted by many factors such as access patterns and traffic pressure that the contribution of the location of a vault is negligible. According to these three insights, we deduce that important NoC parameters such as request size and routing protocol has more contribution in the latency \emph{within an access pattern} rather that physical parameters such as the location of a vault.

\subsection{High-Contention Vault Histograms Per Latency Interval}
\label{sec:res-hist-vault}

To explore the contribution of vaults to high and low latencies, each row of Figure~\ref{fig:histgram-latency} depicts contributing vaults for each latency interval and illustrates the histogram of them. The intensity of the color of a rectangle shows the normalized value of the number of that particular appearance of vault in that latency interval against the maximum number of accesses in that row. Figures~\ref{fig:histgram-latency}a, b, c, and d, show colormaps for various request sizes of 16, 32, 64, and 128\,B, respectively. In Figure~\ref{fig:histgram-latency}a, we observe that for gaining the lowest latency (i.e., lowest row), we should avoid accessing vault numbers 9 to 12. In fact, Figure~\ref{fig:histgram-latency} provides a guide for avoiding certain vaults that incur high latencies, but it will not guarantee particular access latencies for a specific vault (similar to last subsection). For instance, based on Figure~\ref{fig:histgram-latency}c, vault number 2 has the highest contribution to the lowest latency interval, and it similarly has a high contribution for the highest latency. Therefore, the conclusion that accessing only vault number 2 will guarantee the lowest latency is not correct. However, in the same figure, the chance of incurring lower latency increases by avoiding vaults numbers 9 to 12. Even though we cannot reach a unanimous conclusion about the latency of each vault and the hierarchy of NoC in the HMC, which we discussed its reasoning in last subsection, we can conclude that the effects of NoC and vault interactions are not trivial.

Based on the observations mentioned in last paragraph, we interpret that vaults almost equally contribute in high and low latencies. Such behavior suggests two notions to the user or designer of such packet-switched memory: (\romannum{1}) Since lowest latency is obtainable from any vaults, a user may map the memory footprint of an application to optimize other important aspects of accessing these memories, such as access pattern, or request size. In other words, the independency of latency to the physical layout eases the memory mapping constraints. (\romannum{2}) A desirable level of performance to an application can be guaranteed by only understanding and following the lowest and highest resulting latency in any access pattern. Note that the uniformity of vault contributions in latency will be sustainable even in a hierarchical connection of many stacks in another interconnection network for creating a large-scale memory. This is because each stack in this new network would have a similar characteristics.    

\subsection{Requested and Response Bandwidth Analysis}
\label{sec:res-req-res-bw}

\begin{figure}[t]
\centering
\vspace{-0.0in}
\includegraphics[width=1\linewidth]{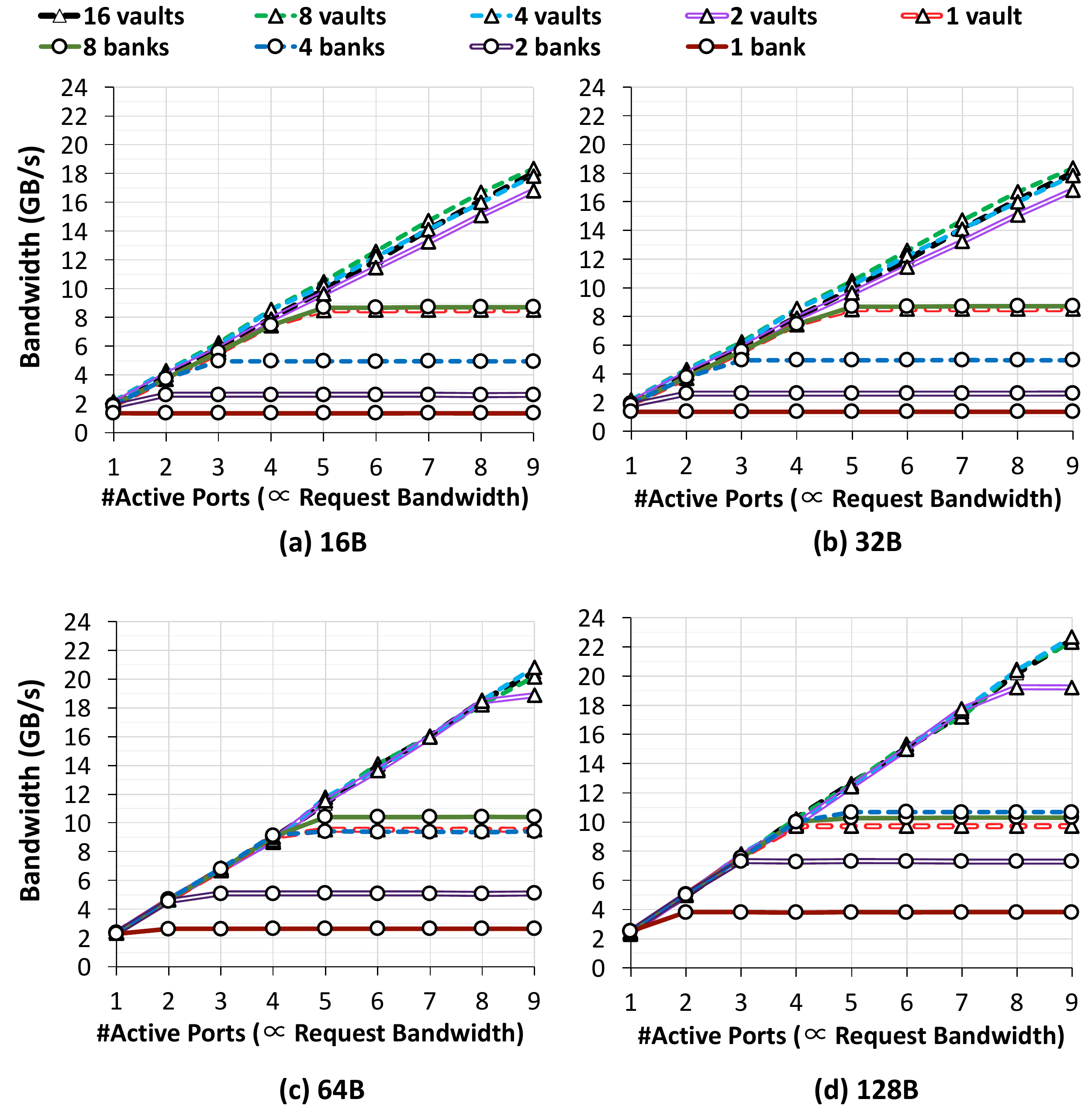}
\vspace{-0.0in}
\captionsetup{singlelinecheck=on,aboveskip=-6pt,belowskip=-12pt}
\caption{Relationships between number of active ports and bandwidth for various request sizes.}
\label{fig:req-res-bw}
\vspace{-0.0in}
\end{figure}

\begin{figure}[t]
\centering
\vspace{-0.0in}
\includegraphics[width=1\linewidth]{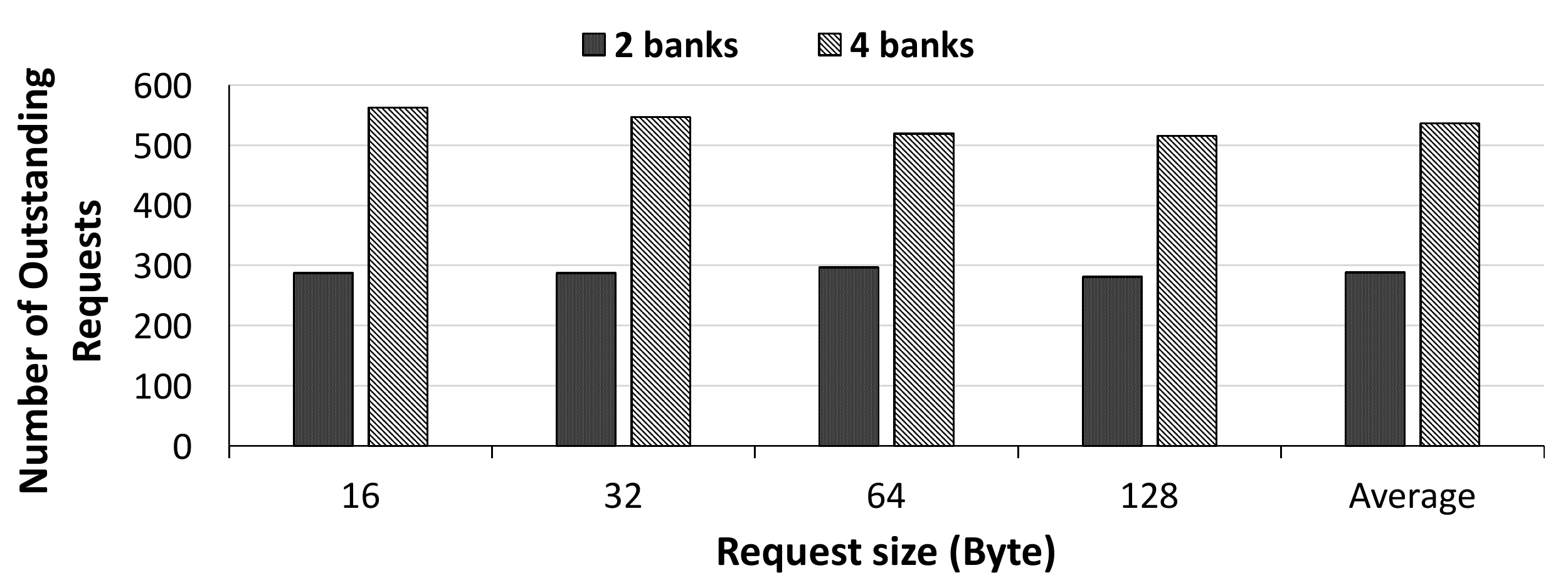}
\vspace{-0.0in}
\captionsetup{singlelinecheck=on,aboveskip=-8pt,belowskip=-10pt}
\caption{Number of estimated outstanding requests in two- and four-bank access patterns.}
\label{fig:fig12}
\vspace{-0.0in}
\end{figure}

To further investigate potential networking bottlenecks and bandwidth of the HMC, we use the GUPS implementation to tune request rate by changing the number of active ports from one to nine ports. The number of active ports is a proxy for the requested bandwidth because it has a direct relationship with the number of issued requests with the GUPS implementation. Figure~\ref{fig:req-res-bw} presents the relationship between the number of active ports and the response bandwidth for various request sizes. In this figure, sloped lines determine access patterns in which no bottleneck occurs. In contrast, flat lines depict access patterns in which a bottleneck (e.g., vault bandwidth limitation) exists. As a recent work about HMC characterization also mentioned~\cite{had:asg17}, the factor that limits the bandwidth utilization can be related to packet-switched network, such as the limited size of queues in the vault controller or DRAM layers. We  analyze the reasons of saturation points by taking a deeper look at a vault controller, which is basically a stationary system, receiving requests with an arrival rate. Based on Little's law, in such systems, the average number of outstanding requests equals to the arrival rate multiplied by the average time a request spends in the system. To calculate the number of outstanding requests based on the numbers represented in Figure~\ref{fig:req-res-bw}, we measure the latency at saturated points and multiply them by input rates, and then divide the result by request size. The result of calculation, illustrated in Figure~\ref{fig:fig12} indicates that regardless of request size, the maximum number of requests is 288 for two banks and 535 for four banks, in average. Moreover, the linear relationship between number of outstanding requests and number of banks suggests that a vault controller dedicates one queue for each bank or for each DRAM layer.

As discussed in Section~\ref{sec:res-high}, we observe that accessing eight banks within a vault saturates the internal 10\,GB/s bandwidth of a vault for request sizes of 16 and 32\,B. In addition, for 64 and 128\,B request sizes, accessing four banks saturates the internal bandwidth of a vault. Thus, within a vault, depending on the size of requests, increasing BLP to more than eight or four banks will not provide higher bandwidth. In fact, as Figure~\ref{fig:address-nocs} presents, for accessing a 4\,KB OS page in the HMC, requests are first spread over vaults and then banks. Therefore, accessing a single page in this configuration naturally avoids this bottleneck. We can extend this insight to more than one OS pages sequentially allocated in the address space. For instance, accessing to more than four sequentially allocated OS pages would invoke the bottleneck of the internal bandwidth of vault. In brief, for effectively utilizing the limited bandwidth of vaults within the HMC, application access patterns must be operated for increasing vault-level parallelism and then bank-level parallelism.

Compared to traditional DRAM memories, HMC supplies a higher amount of bandwidth and concurrency that is due to the high number of vaults and independent vault controllers. Figure~\ref{fig:req-res-bw}d exhibits this point by showing that for request sizes of 128\,B, distributed access patterns to more than two vaults quickly reach the bottleneck of the external bandwidth of two links. This is a limitation of our particular HMC infrastructure (two half links from the FPGA to the HMC), as the number and width of HMC links can be increased as can the speed and efficiency of the FPGA infrastructure (i.e., HMC controller and associated firmware). Since HMC uses bi-directional links, issuing only read requests results in an asymmetric usage of the available bandwidth. In other words, read requests only fully utilize response bandwidth, and write requests only fully utilize request bandwidth. Previous studies~\cite{ros:pau2012, sch:fro2016} have investigated this asymmetry, and proposed issuing a mix of read and write requests to address it. In addition to optimizing access patterns, applications should also balance the ratio of read and write requests for effectively utilizing bi-directional bandwidth of stacked memory networks.

\section{Related Work}
\label{related}
Previous works have characterized the HMC~\cite{sch:fro2016, ros:pau2012, gok:llo15, ibr:fat2016}, from which Schmidt et al.~\cite{sch:fro2016} agreed with our measured bandwidth and latency. Another work, \cite{had:asg17}, using the AC-510 accelerator board, characterized bandwidth of HMC and its relationship with temperature, power, and cooling power. They deconstructed the contributing factors to the latency, but they focused more on power and temperature. Although these studies have explored emulated HMC and earlier HMC prototype chips, they have not studied the performance impacts of the internal NoC on the performance and QoS of the HMC, and in general the impact of packet-switched networks on the performance of 3D-stacked memories.
 
Other recent studies have focused on designing an efficient NoC for HMC. Zhan et al~\cite{zha:jia2016} proposed solving issues that show up in a NoC coupled with HMC, such as traffic congestion, uncoordinated internal and external networks, and high power consumption by co-optimizing networks both inside each HMC and between cubes. Their proposed unified memory network architecture reuses the internal network as a router for the external network, which allows for remote access bypassing while also providing high bandwidth for local accesses. The authors also proposed reducing communication loads and using power gating to further decrease power consumption for an overall 75.1\% reduction in memory access latency and a 22.1\% reduction in energy consumption.

Azarkhish et al~\cite{aza:erf2017} proposed a low latency AXI-compatible interconnect, which provides the required bandwidth for HMC infrastructure so that it supports near memory computation. Their simulation results show that the main bottleneck for delivered bandwidth is the timing of DRAM layers and TSVs. Also, their analysis on PIM traffic with increased requesting bandwidth on the main links showed that when the host demands less than 120 GB/s no saturation occurs. 
In another work, Fujiki~\cite{fuj:dai2016}, et. al proposed a scalable low-latency network by using a random topology based on communication path length, using deadlock-free routing, and memory-mapping in granularity of a page size. Their full-system simulation models shows that this method reduces cycles by 6.6\%, on average, and that random networks with universal memory access out-perform non-random, localized networks.

\section{Conclusion}
\label{sec:con}

In this paper, we evaluate the internal NoC of the HMC, a real-world prototype of a NoC-based, 3D-stacked memory. From our experiments, we can provide the following insights into the effects of the internal NoC of the HMC on the performance of systems and applications.

\begin{itemize}
%
\item 
Large and small request sizes for packets provide a trade-off between effective bandwidth and latency as a result of buffering, packetization, and reordering overheads. In contrast with traditional DDRx systems, this trade-off enables tuning memory accesses to optimize either bandwidth or latency.
(Section~\ref{sec:res-high}, \ref{sec:res-hist-latency}, and \ref{sec:res-req-res-bw})
%
%
\item 
As future memories become denser with more links and vaults, queuing delays will become a serious concern for packet-based memories, such as the HMC. Effective solutions should focus on (\romannum{1}) optimizing queuing on the host controller side and at vault controllers or (\romannum{2}) distributing accesses to improve parallelism, such as BLP.
(Section~\ref{sec:res-low} and \ref{sec:res-qos})
%
%
\item 
The internal NoC complicates QoS for memory accesses because of meaningful variations in latency even within an access pattern. On the other hand, it creates opportunities such as (\romannum{1}) smaller packets are ensured to have improved QoS at a cost of reduced bandwidth or (\romannum{2}) high-priority traffics can be mapped to access their private vaults.    
(Section~\ref{sec:res-qos}, \ref{sec:res-hist-latency}, and \ref{sec:res-hist-vault})
%
%
\item 
Limited bandwidth within a vault means that mapping accesses across vaults then banks is key to achieve better bandwidth utilization and lower latency.
(Section~\ref{sec:res-high} and \ref{sec:res-req-res-bw})
%
%
\item 
The packet-based protocol creates an asymmetric bi-directional bandwidth environment that applications should be aware of and optimize for the proper mix of reads and writes for effectively utilizing external bandwidth.
(Section~\ref{sec:res-high} and \ref{sec:res-req-res-bw})
%
%
\item 
Finally, the exact latency of a vault is impacted by many factors such as access patterns and traffic conditions that the latency contribution of the physical location of a vault is negligible within an access pattern. This insight reduces complexity and constraints of optimization and mapping techniques.
(Section~\ref{sec:res-hist-latency} and \ref{sec:res-hist-vault})
\end{itemize}

\IEEEtriggeratref{6}
\IEEEtriggercmd{\enlargethispage{-3.5in}}

\bibliographystyle{IEEEtran.bst}
\bibliography{short}

\clearpage

\end{document}